\documentclass[fleqn,usenatbib]{mnras}

\usepackage{natbib}
\bibpunct{(}{)}{;}{a}{}{,}
\usepackage{graphicx}
\usepackage{txfonts}
\usepackage{xcolor}
\usepackage{multirow}

\def\inte{{\em INTEGRAL}}
\def\xmm{{\em XMM-Newton}}
\def\heao{{\em HEAO 1}}

\def\chan{{\em Chandra}}
\def\asca{{\em ASCA}}
\def\beppo{{\em BeppoSAX}}
\def\rxte{{\em RXTE}}
\def\swift{{\em Swift}}
\def\rosat{{\em ROSAT}}

\def\suzaku{{\em Suzaku}}
\def\nustar{{\em NuSTAR}}
\def\ginga{{\em Ginga}}
\def\maxi{{\em MAXI}}

\def\ferg{\mathrm{erg\,s^{-1}\,cm^{-2}}}

\def \srca {Sct\,X$-$1}
\def \srcb {IGR\,J17329$-$2731}
\def \srcc {4U\,1700$+$24}

 \definecolor{arancio}{rgb}{1,0.5,0}
 \definecolor{viola}{rgb}{0.7,0,1}
 \definecolor{verde}{rgb}{0.2,0.7,0.7}
\definecolor{cobalt}{rgb}{0.0, 0.28, 0.67}
\definecolor{airforceblue}{rgb}{0.36, 0.54, 0.66}
\definecolor{ballblue}{rgb}{0.13, 0.67, 0.8}
\definecolor{battleshipgrey}{rgb}{0.52, 0.52, 0.51}
\definecolor{darkgreen}{rgb}{0.0, 0.2, 0.13}

\title[Symbiotic X-ray binaries]{The Symbiotic X-ray binaries Sct\,X-1, 4U\,1700+24 and IGR\,J17329-2731}
\author[E. Bozzo et al.]{
E.\ Bozzo,$^{1}$,$^{2}$\thanks{E-mail: enrico.bozzo@unige.ch}
P.\ Romano,$^{3}$
C.\ Ferrigno,$^{1}$
and L.\ Oskinova,$^{4}$
\\
$^{1}$Department of Astronomy, University of Geneva, Chemin d'Ecogia 16, CH-1290 Versoix, Switzerland\\
$^{2}$INAF-OAR, Via Frascati, 33, 00078 Monte Porzio Catone, Rome, Italy\\
$^{3}$INAF, Osservatorio Astronomico di Brera, Via E.\ Bianchi 46, I-23807, Merate, Italy\\
$^{4}$Institut f\"ur Physik und Astronomie, Universit\"at Potsdam, Karl-Liebknecht-Strasse 24/25, 14476 Potsdam, Germany
}

\date{}

\begin{document}
\label{firstpage}
\pagerange{\pageref{firstpage}--\pageref{lastpage}}
\maketitle

\begin{abstract}
Symbiotic X-ray binaries are systems hosting a neutron star accreting form the wind of a late type companion. These are rare objects and so far only a handful of them are known. One of the most puzzling aspects of the symbiotic X-ray binaries is the possibility that they contain strongly magnetized neutron stars. These are expected to be evolutionary much younger compared to their evolved companions and could thus be formed through the (yet poorly known) accretion induced collapse of a white dwarf. In this paper, we perform a broad-band X-ray and soft $\gamma$-ray spectroscopy of two known symbiotic binaries, Sct\,X$-$1 and 4U\,1700$+$24, looking for the presence of cyclotron scattering features that could confirm the presence of strongly magnetized NSs. We exploited available \chan, \swift, and \nustar\ data. We find no evidence of cyclotron resonant scattering features (CRSFs) in the case of Sct\,X$-$1 but in the case of 4U\,1700$+$24 we suggest the presence of a possible CRSF at $\sim$16~keV and its first harmonic at $\sim$31~keV, although we could not exclude alternative spectral models for the broad-band fit. If confirmed by future observations, 4U\,1700$+$24 could be the second symbiotic X-ray binary with a highly magnetized accretor. We also report about our long-term monitoring of the last discovered symbiotic X-ray binary IGR\,J17329$-$2731 performed with \swift/XRT. The monitoring revealed that, as predicted, in 2017 this object became a persistent and variable source, showing X-ray flares lasting for a few days and intriguing obscuration events that are interpreted in the context of clumpy wind accretion.
\end{abstract}


\section{Introduction}
\label{sec:intro}

Symbiotic X-ray binaries (SyXBs) are rare low mass X-ray binaries (LMXBs) hosting a red giant and (in most cases) a neutron star (NS) accreting from the slow wind of its companion. Only five\footnote{We take into account here the fact that 3A\,1954+319 has been recently reclassified as a classical supergiant X-ray binary \citep{hinkle20,bozzo22}.} SyXBs are known so far \citep{masetti06, masetti07, masetti07b, nucita07, corbet08, nespoli08, bozzo13, bozzo18}.

Although these sources are formally part of the LMXBs class due to the low mass of the red giant donor star, their behaviour in the X-ray domain strongly resembles that of the high mass X-ray binaries (HMXBs). SyXBs show long pulse periods, ranging from hundred of seconds to hours, and
display a high pulsed fraction that can be as large as 30-50\,\%. They are characterized by the longest orbital periods among the known LMXBs (several tens to thousand days) and a prominent variability in the X-ray domain by a factor of $\sim$10-20 typical of wind-fed
HMXBs. These properties strongly suggest the presence of a highly magnetized NS (at least $>$10$^{12}$~G) in the SyXBs.

The evolutionary calculations available so far \citep{postnov11, lu12, kuranov15} assume \emph{ a priori} that the progenitor of a SyXB is a binary system initially hosting a strongly magnetized NS and a low mass main sequence star ($\lesssim$2~M$_{\odot}$) in a wide orbit (hundreds of days). In the early stages of  evolution, the secondary star is on the first giant branch but does not fill its Roche lobe, leading to a negligible mass transfer toward the NS and thus to negligible high-energy emission. When this is no longer the case, as the magnetic field of the NS is assumed not to have decayed and its rotational velocity is still high, the system enters in a so-called ``propeller'' phase. Inflowing material from the donor star is ejected from the vicinity of the NS and accretion is still largely inhibited. This occurs at the expense of the rotational energy, inducing a rapid increase of the NS spin period up to $\gtrsim$10\,000~s \citep{bozzo08,shakura12}. As little to no accretion is taking place in this stage, the system is hardly detectable in the X-ray domain. The situation changes when the NS has slowed down sufficiently to reduce the centrifugal force at the magnetospheric boundary and allows some accretion to take place, finally shining as a SyXB. In this phase, accretion is still likely to take place directly from the stellar wind. As this is endowed with little to no angular momentum, the spin period of the NS is still expected to increase because of the effect of the friction between the relatively large magnetosphere and the surrounding dense environment. In the following evolutionary phase, the velocity of the stellar wind decreases to a level at which the formation of an accretion disk around the compact object is inevitable \citep{wang81}. Accretion in this phase leads to rapid decrease of the NS spin period and the system finally resembles a common LMXB with a fast spinning compact object \citep[see Fig.~1 of][]{lu12}.

One of the most puzzling aspects of SyXBs concerns the presence of a strongly magnetized NS in a system that has to be at least several Gyr old to allow the donor star to evolve to the red giant phase. The measurement of the NS magnetic field is usually obtained through the detection of cyclotron resonant scattering features (CRSFs). These are absorption features in the X-ray spectra of NS binaries that were first discovered in Her X-1 \citep{trumper77,trumper78} and later identified in several other systems \citep{heindl04,schonherr2007}. Long-standing theoretical developments have shown that the NS magnetic field should decay with time especially in the presence of accretion, and this is the main reason why the bulk of the LMXBs host weakly magnetized NSs \citep{tauris15}. An alternative possibility put forward is that the NS formed much later in the evolution of the SyXB due to the accretion induced collapse of a white dwarf accreting from the stellar wind of the red giant. The white dwarf would have an age compatible with that of the red giant companion, while the NS forms more recently after a long phase of accretion. This allows the NS to preserve its high magnetic field up to the point in time at which the system shines as a SyXB. So far, the accretion induced collapse of a white dwarf remains a relatively poorly known process and it is a matter of debate if a strongly magnetized NS (at the level of $\gtrsim$10$^{12}$~G) can be produced by conservation of a sufficiently intense white dwarf magnetic flux during the collapse \citep[see, e.g.,][and references therein]{tauris15}.

Only in the case of the SyXB \srcb{} the presence of a strongly magnetized NS could be firmly established thanks to the the detection of a CRSF at $\sim$21\,keV, providing an estimate of the compact magnetic field strength as high as 2.4$\times$10$^{12}$~G \citep{bozzo18}. This discovery called for further observations of \srcb,\ as well as of other SyXBs, to consolidate our understanding on the formation channels of these systems.

In this paper, we pursue the search for CRSFs in SyXBs through the broad-band spectral analysis of the X-ray emission from two known sources in this class, \srca\ and \srcc.\ For the first source, we report on our simultaneous observation with \chan\ and \nustar.\ For \srcc,\ we exploit a public but yet unpublished \nustar\ observation, as well as previously performed broad-band observations with the XRT and BAT on-board \swift\ during three  outbursts of the source that occurred in 2014 and 2015. We additionally report in this paper on the outcomes of our long-term observational campaign with \swift/XRT on the most recently discovered SyXB \srcb,\ covering up to nine months after its first detection in the X-rays (August 2017).

\section{ \srca}
\label{sec:sctx1}

\srca{}was discovered in 1974 \citep{hill74} and subsequently observed with a variety of X-ray facilities. The system is known to host a 111\,s spinning NS and is characterized by a remarkably high local absorption column density reaching up to $\sim10^{23}$\,cm$^{-2}$. It also showed a prominent variability (from $<$0.3 to 20\,mCrab, corresponding to 5$\times10^{-12}$ to 3$\times10^{-10}$\,erg\,cm$^2$\,s$^{-1}$) on a large range of timescales and the presence of a neutral iron line emission that was detected in its X-ray spectrum already back in the '90s \citep[the \heao\ satellite detected \srca\ up to $\sim$100\,keV; see][and references therein]{koyama91}. The best determined source position to date was obtained through an \xmm\ observation reported by \citet{kaplan07}. These authors could use the improved source localization (1~arcsec) to identify the optical counter-part of the X-ray source and proposed that the NS in \srca\ is coupled to a late type giant about 4\,kpc away from us. This led to the classification of \srca{} as a SyXB in our Galaxy. The \xmm\ data revealed also that the source had continuously spun down from the late '90s to 2004, and the best measured pulse period from \xmm\ was 112.86$\pm$0.08\,s. \citet{kaplan07} further found that since the late '90s the source had decreased in flux substantially, reaching at the time of the \xmm\ observation a flux of 0.4~\,Crab (corresponding to 1.2$\times10^{-11}$~erg~cm$^2$~s$^{-1}$ in the 0.5--10\,keV energy band). The spectrum measured by the EPIC cameras on-board \xmm\ proved to be significantly harder compared to previous measures with \heao\ and \ginga\ (the power-law photon index decreased from $\sim$2.0 to $\sim$1.5), and the absorption column density was shown to have decreased by a factor of few \citep[from 2--4$\times10^{23}$\,cm$^{-2}$ to $8\times 10^{22}$\,cm$^{-2}$; see][]{heao,ginga}. 
No additional X-ray observations were performed toward \srca\ between 2004 and 2014. More recently, \citet{kishalay22} reported on the first broad-band IR spectroscopic investigation of the source and re-classified the companion as a M8-9 III type O-rich Mira donor star, rather than a red giant. This makes \srca\ a peculiar member of the SyXBs and the first known NS binary with a Mira companion.

In this paper, we report on an observational campaign aimed at \srca\ (almost) simultaneously with \chan\ and \nustar.\ We also exploit two yet unpublished archival \suzaku\ observations pointed at the source.

\srca\ was observed with \chan\ on 2020 November 9 at 01:50 (UT) for a total exposure time of 28.7\,ks (PI: E. Bozzo). The observation was carried out with the ACIS-S in Faint mode (timed exposure). We processed the \chan\ data with standard techniques\footnote{See \href{https://cxc.cfa.harvard.edu/ciao/threads}{https://cxc.cfa.harvard.edu/ciao/threads}.} using CIAO v4.3 and the latest CALDB available at the moment of writing (v.0460; released on 2021 September 23). Only one source was detected in the ACIS image extracted in the 0.5--10\,keV and the  best determined position using the CIAO (v4.3) tool {\sc celldetect} is at RA$=$278.85755 and DEC$=-$7.61408 with an associated uncertainty at 90\,\% c.l. of 0.6~arcsec. This position is fully consistent with that previously reported by \citet{kaplan07,kishalay22} and within 0.3~arcsec from the selected counterpart of \srca{} (2MASS J18352582$-$0736501). This confirms the classification of \srca\ with the Galactic SyXB.

Using the CIAO tool {\sc dmextract}, we determined that only about 170 effective counts were recorded from the source and thus we did not attempt any timing investigation, limiting the analysis to the sole extraction of a time-averaged spectrum (rebinned to have at least 25   counts per energy bin). The corrected source spectrum was extracted using the CIAO tool {\sc specextract} and could be well fit ($\chi^2$/d.o.f.$=6.1/7$) with a simple absorbed powerlaw model ({\sc tbabs*pow} in {\sc Xspec}). We adopted the {\em wilm} abundances \citep{wilms00} and {\em vern} cross sections \citep{vern96}. We measured an absorption column density of $N_{\rm H}$=(5.0$_{-1.8}^{+2.5})\times10^{22}$~cm$^{-2}$ and a photon index of $\Gamma=1.8_{-0.7}^{+0.9}$. The estimated 0.5--10~keV X-ray flux was 2.1$\times10^{-13}$~erg~cm$^{-2}$~s$^{-1}$. These results are fully consistent with those reported by \citet{kishalay22}.
\begin{figure}
 \centering
 \includegraphics[width=5.9cm,angle=-90]{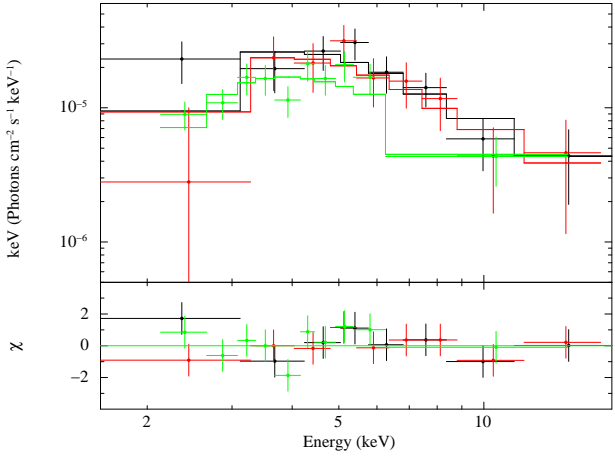}
  \caption{Broad-band (unfolded) X-ray spectrum of \srca\ combining the \chan\ ACIS-S (green) with the \nustar\ FPMA (black) and FPMB (red) data. The best fit is obtained with a model comprising an absorbed powerlaw. The residuals from the fit are shown in the bottom panel.}
  \label{fig:broadband}
\end{figure}

\nustar\ observed \srca\ starting on 2020 November 8 at 18:16 for a total exposure time of 32.5~ks (PI: E. Bozzo). We reduced the data-set using standard  techniques\footnote{See
\href{https://heasarc.gsfc.nasa.gov/docs/nustar/analysis/nustar\_swguide.pdf}{https://heasarc.gsfc.nasa.gov/docs/nustar/analysis/nustar\_swguide.pdf} } and the caldb version released in 2021 December 2. We first extracted both the FPMA and FPMB images and noted that a single faint source was detected at a location compatible with the \chan\ position of \srca{} reported above. Given the faintness of the source (about 320 counts were collected by both the FPMs), it was not possible to carry out a study of the source lightcurve or perform any timing analysis. We thus extracted only the time averaged spectra (rebinned to have at least 25 counts per energy bin) and fit the data from both FPMs together. The source spectrum could be well described by an absorbed powerlaw model ($\chi^2$/d.o.f.$=9.0/13$). We measured $N_{\rm H}$=(14.2$_{-10.8}^{+16.3})\times10^{22}$~cm$^{-2}$ and $\Gamma=3.8_{-0.9}^{+1.2}$. The estimated 3--20\,keV X-ray flux was 2.5$\times10^{-13}$~erg~cm$^{-2}$~s$^{-1}$. A normalization constant was included in the fit to take into account cross-calibrations between the FPMA and FPMB but it turned out to be fully compatible with unity.

We also fit simultaneously the \chan\ and \nustar\ data. In this case, we obtained $N_{\rm H}$=(8.6$_{-3.0}^{+3.8})\times10^{22}$~cm$^{-2}$ and $\Gamma=2.9_{-0.5}^{+0.6}$ ($\chi^2$/d.o.f.$=17.7/22$). The normalization constant introduced to take into account cross-calibrations between \chan\ and \nustar\ was measured at 0.6$\pm$0.2 (fixing the \nustar\ constant to unity for reference). We show the results of the fit to the broad-band spectrum of \srca\ in Fig.~\ref{fig:broadband}.

We noted that two \suzaku\ observations of \srca{} were also carried out in 2014 October 10 at 12:43 and in 2014 October 22 at 02:45 (UT). The exposure time available for the XIS instrument was of 15.2\,ks and 38.6\,ks, respectively. We obtained from the {\sc Heasarc} archive the pre-processed XIS data and verified that no source was detected at a position consistent with \srca. We adopted the {\sc ximage} tool and the command {\sc uplimit} available within {\sc Heasoft} v.6.29c to estimate an upper limit on the source X-ray flux \citep[see][and references therein]{bozzo12}. From the shorter observation, we obtained a 3$\sigma$ upper limit on the source count-rate of 0.22\,cts\,s$^{-1}$, while the longer observation returned an upper limit of 0.15\,cts\,s$^{-1}$. By using the online tool {\sc webpimms}\footnote{\href{https://heasarc.gsfc.nasa.gov/cgi-bin/Tools/w3pimms/w3pimms.pl}{https://heasarc.gsfc.nasa.gov/cgi-bin/Tools/w3pimms/w3pimms.pl}} and the spectral model measured from the \chan\ $+$ \nustar\ data, we converted the above values into a 3\,$\sigma$ upper limit on the source 0.5--10~keV X-ray flux of 8.6$\times10^{-13}$~erg~cm$^{-2}$~s$^{-1}$  and 5.6$\times10^{-13}$~erg~cm$^{-2}$~s$^{-1}$, respectively.

\section{\srcc}
\label{sec:4u1700}

\srcc\ was discovered in the '70s and it is known since then to be a relatively bright persistent X-ray source \citep[see, e.g.,][and references therein]{garcia83,nucita14,hinkle19}. The source is known to host an accreting NS which pulsations have been so far elusive most likely due to the pole-on direction through which the system is being observed. The source shows a remarkable variability in X-rays, achieving a total dynamic range of at least $\sim$200, considering also that the source has been detected undergoing short outbursts a few times by both \swift\ and \maxi\  \citep{kennea14,fukushima14,burrows14,burrows15}. The M giant companion is relatively well studied and it has been recently reported though long-term optical and near-IR observations the discovery of its pulsating period (about 420~days), as well as the orbital period of the system measured at $\sim$12.0~years \citep{hinkle19}. This is by far the longest known orbital period for a SyXB and the detailed modeling of the optical data provided support in favor of the previous finding that the system is observed pole-on, with an estimated orbital inclination of 11.3$\pm$0.4$^{\circ}$. The measurement of the distance to the source has been also recently improved thanks to the Gaia data at 0.536$\pm$0.009~kpc \citep{arnason21}.

In the soft X-ray domain ($\lesssim$10~keV), the spectral energy distribution of the source is usually described by using a model comprising a thermal black-body emission dominating below $\sim$2~keV and a comptonization component \citep[see, e.g.,][and references therein]{nucita14}. A red-shifted O VIII Ly-$\alpha$ transition line has been detected multiple times by using the RGSs on-board \xmm\ and so far ascribed to either the re-organization of the X-ray emitting material close to the NS magnetic poles or to the presence of a uni-polar jet of matter emitted by the NS with velocities of the order of few 1000~km~s$^{-1}$. A study of the broad-band X-ray emission from \srcc\ was carried out in the past by \citet{masetti02} using a combination of \rosat,\ \asca,\ \rxte,\ and \beppo\ data, as well as by \citet{nagae08} using \suzaku\ data. No evidence of a cyclotron line was found, with the broad-band spectra of the source being well described by the same combination of components mentioned above.

In this paper, we report on a yet unpublished \nustar\ observation of \srcc\ carried out from 2014 October 2 at 10:26 (UT) to October 3 at 15:21 for a net
exposure time of 51.6~ks over 102~ks of observation. We made use of data from both the FPMA and FPMB in the energy range 3--79~keV, processing the data with the
\texttt{nupipeline} (version 0.4.9) available within the {\sc Heasoft} (version 6.29) and the version 20211020 of the calibration database. The processing
provided us the cleaned event files. We used extraction regions of two arcminutes radius for both source and background products. These region files were used
as input to \texttt{nuproducts} (version 0.3.3) to obtain the source and background lightcurves and spectra. We verified that different reasonable
choices of the location of the background region and of the radius of extraction do not affect our results.

\begin{figure}
    \centering
    \includegraphics[height=\columnwidth,angle=-90]{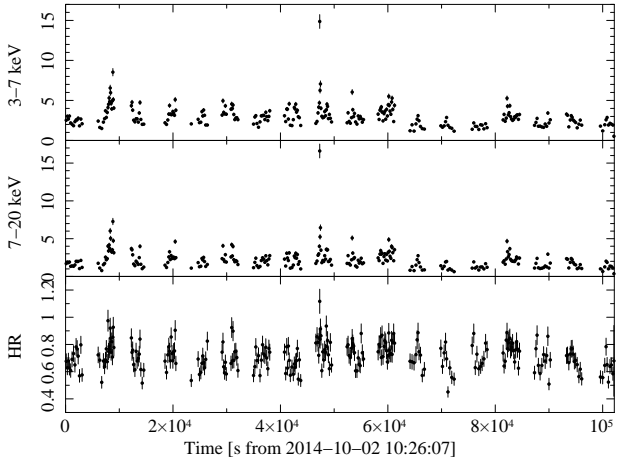}
    \caption{From top to bottom \nustar\ soft (3--7~keV), hard (7--20~keV) lightcurves of \srcc\ and their ratio (HR).}
    \label{fig:u1700_nustar_lc}
\end{figure}

The source lightcurve was extracted with bins of one second and adaptively rebinned to reach a signal to noise ratio of at least 25 in each bin \citep[see][for details on the adaptive hardness ratio rebinning employed in several of our papers]{bozzo13}.
As it can be appreciated in Fig.~\ref{fig:u1700_nustar_lc}, the source lightcurve shows a moderate variability that is not accompanied by significant changes in the spectral properties (the hardness ratio, HR, remained virtually constant across the entire observation). This was confirmed by the usage of a Bayesian block analysis \citep[the same exploited in a number of our previous papers; see, e.g.,][]{ferrigno20} which thus convinced us to extract a single spectrum using the entire exposure time available to investigate the properties of the broad-band X-ray emission from \srcc.\

The spectrum in the 3--70\,keV range was rebinned using the optimal grouping method described by \citet{Kaastra} and the
Cash statistics (\texttt{C-stat} in \textsc{Xspec}) was adopted to evaluate the goodness of all fits.
We first tested a simple phenomenological model to describe the source spectrum, comprising a cut-off power-law component (\texttt(highecut*pegpwrlw) in \textsc{xspec}) attenuated by neutral absorption at the lower energies (\texttt(TBabs) in \textsc{xspec}). We adopted for the absorption component Vermmer cross sections \citep{verner95aa} and ``Wilm'' abundances \citep{wilms00}. The best-fit parameters were determined using the modified
Levenberg-Marquardt algorithm based on the CURFIT routine from Bevington, while uncertainties on the parameters were obtained by running a Monte Carlo Markov Chain (MCMC) with the Goodman-Weare sampling algorithm. We used 60 walkers, a burn-out of 6000 steps, and a chain length of 36\,000. Priors were uniform for the slope and uniform in logarithmic scale for all other parameters. We set parameter limits wide enough not to influence the posteriors.
To asses the goodness of the fit, we sampled 1\,000 parameter sets derived from the MCMC, simulated a spectrum based on that model with the same
exposure and background of the measured one, and performed a fit. We compared the C-stat of the simulated set to the one derived from the data and found that all simulations yield a better fit statistic. We thus concluded that the model had a probability less than 0.1\% to be statistically acceptable.

\begin{figure}
    \centering
    \includegraphics[width=\columnwidth,angle=0]{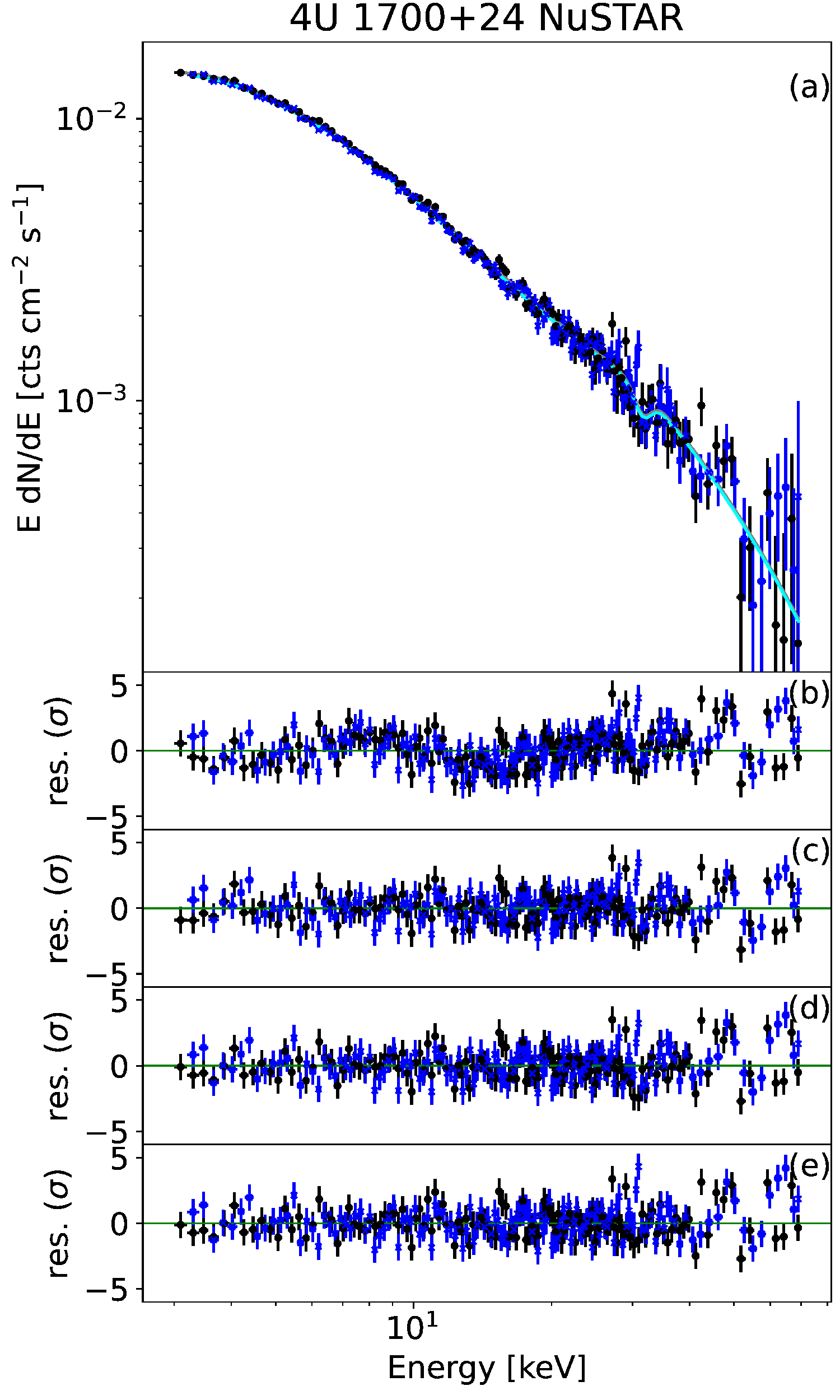}
    \caption{Panel (a): unfolded FPMA (black) and FPMB (blue) spectra of \srcc. The gray and cyan lines are the best-fit model for FPMA and FPMB, respectively, obtained by adopting the model \texttt{TBabs*gabs*gabs*highecut*pegpwrlw}. Panel (b): residuals from the fit with the model \texttt{TBabs*highecut*pegpwrlw}. Panel (c): residuals from the fit with the model \texttt{TBabs*highecut*(bbodyrad+pegpwrlw)}. Panel (d): residuals from the fit with the model \texttt{TBabs*gabs*highecut*pegpwrlw}. Panel (e): residuals from the fit with the model \texttt{TBabs*gabs*gabs*highecut*pegpwrlw}. FPMA and FPMB data were rebinned here only for plotting purposes.}
    \label{fig:u1700_nustar_spec}
\end{figure}

Driven by the visual inspection of the residuals from the above fit, we tested the improvement of the results by adding to the simple phenomenological model either a black-body or two broad Gaussian absorption lines with centroid energies fixed to be one the double of the other. These two more complete models gave equivalently acceptable results. The black-body radius turned out in the fit to have a radius compatible with that of an hot spot on the NS surface, providing a 17\% contribution to the total 1--10\,keV unabsorbed flux. In the model featuring absorption Gaussian lines, the width of the fundamental feature is of 8 or 5~keV depending if we include or not the first harmonic in the fit. The latter turned out to be only marginally significant, but the different panels reported in Fig~\ref{fig:u1700_nustar_spec} shows how the introduction of this second feature is able to treat the otherwise evident residuals around 30\,keV. All results of the above fits are summarized in Table~\ref{tab:4u_spectral}. For completeness, we also tested alternative models for the description of the source continuum emission that are commonly exploited in case of strongly magnetized accreting NSs, as the Fermi-Dirac cutoff or the NPEX. None of these provide significant improvements over the phenomenological model and thus are not further discussed.
\begin{table}
    \begin{center}
    \label{tab:4u_spectral}
    \caption{Spectral fit results for the average \nustar\ spectrum of \srcc. Uncertainties are at 68\% confidence level derived from the 16 and 84\% quantiles of the MCMC. }
    \begin{tabular}{lr@{}ll}
        \hline
        \hline
        \multicolumn{4}{c}{\texttt{Tbabs*highecut*(pegpwrlw+bbodyrad)}}\\
        \hline
        $N_\mathrm{H}$ & 1.8 &$\pm$0.6 & $10^{22}$cm$^{-2}$\\
        $E_\mathrm{C}$ & 10.7 &$_{-0.8}^{+0.6}$ & keV \\
        $E_\mathrm{F}$ & 49 &$_{-6}^{+7}$ & keV \\
        $T_\mathrm{bb}$ & 1.43 &$\pm$0.03 & keV\\
        $r$ & 0.86 &$\pm$0.08 & km/10 kpc\\
        $\Gamma$ & 1.96 &$\pm$0.05 & \\
        Flux (1--10\,keV)$^a$ & 167 &$\pm$13 & $10^{-12}$erg\,s$^{-1}$\,cm$^{-2}$ \\
        Simulated fraction$^b$ & 23&\% & \\
        \hline
        \hline
        \multicolumn{4}{c}{\texttt{TBabs*gabs*highecut*pegpwrlw}}\\
        \hline
        $N_\mathrm{H}$ & 4.3 &$_{-0.3}^{+0.4}$ & $10^{22}$cm$^{-2}$\\
        $E_\mathrm{Cyc1}$ & 16.4 &$_{-0.5}^{+0.7}$ & keV \\
        $\sigma_\mathrm{Cyc1}$ & 5.1 &$_{-0.7}^{+1.4}$ &  \\
        $\tau_\mathrm{Cyc1}$ & 1.7 &$_{-0.4}^{+1.0}$ &  \\
        $E_\mathrm{C}$ & 7.5 &$\pm$0.4 & keV\\
        $E_\mathrm{F}$ & 43 &$\pm$4 & keV\\
        $\Gamma$ & 2.07 &$\pm$0.04 & \\
        Flux (1--10\,keV) & 248 &$_{-5}^{+8}$ & $10^{-12}$erg\,s$^{-1}$\,cm$^{-2}$ \\
        Simulated fraction$^b$ & 27&\% & \\
        \hline
        \hline
        \multicolumn{4}{c}{\texttt{TBabs*gabs*gabs*highecut*pegpwrlw}}\\
        \hline
        $N_\mathrm{H}$ & 4.2 &$_{-0.5}^{+0.3}$ & $10^{22}$cm$^{-2}$\\
        $E_\mathrm{Cyc1}^c$ & 16.0 &$_{-0.4}^{+0.6}$ & keV \\
        $\sigma_\mathrm{Cyc1}$ & 8 &$\pm$1 & keV \\
        $\tau_\mathrm{Cyc1}$ & 5 &$_{-2}^{+2}$ &  \\
        $\sigma_\mathrm{Cyc2}$ & 1.6 &$_{-0.9}^{+1.7}$ & keV \\
        $\tau_\mathrm{Cyc2}$ & 0.4 &$_{-0.3}^{+0.5}$ & \\
        $E_\mathrm{C}$ & 7.6 &$\pm$0.4 & keV\\
        $E_\mathrm{F}$ & 35 &$\pm$4 & keV\\
        $\Gamma$ & 2.01 &$_{-0.05}^{+0.04}$ &  \\
        Flux (1--10\,keV) & 259 &$_{-10}^{+13}$ & $10^{-12}$erg\,s$^{-1}$\,cm$^{-2}$ \\
        Simulated fraction$^b$ & 45&\% & \\
        \hline
    \end{tabular}\\
\end{center}
Notes: $^a$ The 1-10~keV flux is relative to the power-law without considering the absorption. $^b$ this is the fraction of spectra simulated starting from a model derived from the MCMC with fit statistic larger than the best fit of the data. $^c$ The energy of the putative harmonic is fixed to be $2\times E_\mathrm{Cyc1}$.

\end{table}

The \swift/BAT Transient Monitor\footnote{\href{https://swift.gsfc.nasa.gov/results/transients/weak/4U1700p24/}{https://swift.gsfc.nasa.gov/results/transients/weak/4U1700p24/}} \citep[][]{Krimm2013TM} shows that \srcc{} was very active during 2014 and early 2015. Indeed, it triggered the BAT three times within four months.
In this paper, we exploit the quasi-simultaneous BAT and XRT data collected during these triggers (see the complete log in Table~\ref{u1700:tab:swift_xrt_log}) in order to perform an additional broad-band spectral analysis of the source beside that made possible by the previously reported \nustar\ data. The \swift\ data of \srcc{} were uniformly processed and analysed using the standard software ({\sc FTOOLS}\footnote{\href{https://heasarc.gsfc.nasa.gov/ftools/ftools_menu.html}{https://heasarc.gsfc.nasa.gov/ftools/ftools\_menu.html.}} v6.29b), calibration (CALDB\footnote{\href{https://heasarc.gsfc.nasa.gov/docs/heasarc/caldb/caldb_intro.html}{https://heasarc.gsfc.nasa.gov/docs/heasarc/caldb/caldb\_intro.html.}}  20210915), and methods. We analysed the BAT event data with the standard BAT software
and created mask-tagged light curves for each trigger in the standard energy bands
(15--25, 25--50, 50--100, 100--350\,keV) rebinned so as to fulfill at least one
of the following conditions: reaching a signal-to-noise ratio (S/N) of 5 or bin length of 10\,s.
Mask-weighted spectra were also extracted from the events collected during the first orbit
of each observation (`first orbit' spectrum)
and during the brightest part of the outburst\footnote{We adopt the T$_{\rm total}$ baseline, that is the time over which all fluence is emitted, as calculated by {\sc battblocks}.} (`bright' spectrum).
An energy-dependent systematic error vector was applied to the data and response matrices
were created with {\sc batdrmgen} using the latest spectral redistribution matrices.
The \swift/XRT data were filtered with the task {\sc xrtpipeline} (v0.13.6). Data below 1~keV were discarded in order to avoid known instrumental residuals affecting the WT mode \footnote{\href{https://www.swift.ac.uk/analysis/xrt/digest_cal.php}{https://www.swift.ac.uk/analysis/xrt/digest\_cal.php.}}. In the following, we analyse the observations corresponding to the different BAT triggers (ObsID 00612974000, 00621278000, and 00623434000), each consisting of three orbits of data. For each trigger, we selected a quasi-simultaneous pair of XRT and BAT spectra. Fits to each pair of spectra were performed using the same models discussed before for the \nustar\ data and including a constant to take into account both the difference of exposure and the non strict simultaneity of the XRT and BAT data.

\begin{figure}
\centering
\hspace{-0.5truecm}
\includegraphics[width=6.2cm,angle=-90]{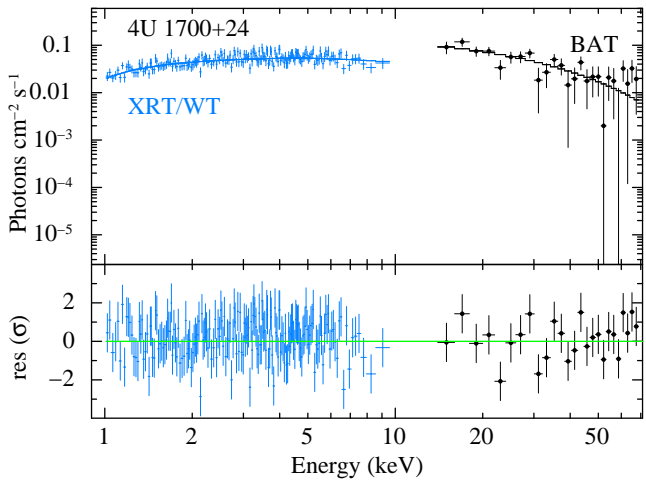}
    \caption{Spectroscopy of the 2014 September 17 outburst of \srcc\ as recorded by \swift.\
   Top panel: unfolded spectra of the nearly-simultaneous XRT/WT data (blue crosses) and BAT  data (black circles) fit with the \texttt{TBabs*highecut*pegpwrlw}  model.
   Bottom panel: residuals from the fit. }
    \label{fig:u1700_xrt_bat_spec_t1}
\end{figure}

\srcc{} triggered the BAT for the first time on 2014 September 17 at 13:30:35.5 UT
(image trigger 612974, T$_0=$MJD 56917.56291043; \citealt[][]{burrows14});
\swift{} immediately slewed to the target so that the narrow field instruments (NFI)
started observing the source at $T_0+$1183\,s (see Table~\ref{u1700:tab:swift_xrt_log}).
During the observation ObsID 00612974000,
the source was significantly detected in the BAT up to $\sim 70$\,keV and
the BAT lightcurve shows only a moderate variability. We checked that, when fit with a simple power-law,
the first orbit spectrum and the bright spectrum (T$_0+524$--$733$\,s)
feature consistent parameters despite a difference by factor of two in flux.
In the XRT, the source showed significant flux variability not corresponding to comparable
variability in its spectral properties (as also noted by \citealt[][]{burrows14}). The
pile-up corrected count rate is seen to vary from a maximum of $\sim 70$\,c\,s$^{-1}$
to a minimum of $\sim 2$\,c\,s$^{-1}$ but the hardness ratio calculated in the
0.3--4\,keV and 4--10\,keV energy bands does not show significant variations.
XRT data collected in the first and third orbit comprised only a few seconds of exposure, with the bulk of the photons collected
during the second orbit.
To maximize the signal-to-noise ratio, we thus considered for our broad-band spectral analysis only the quasi-simultaneous
BAT bright spectrum (T$_0+524$--$733$\,s) and the XRT/WT spectrum from the second orbit (T$_0+5634$--$6188$\,s).
A fit to these data using an absorbed power-law with an an exponential cutoff (\texttt{TBabs*highecut*pegpwrlw}) provided satisfactory results. We report in Table~\ref{u1700:tab:swift_xrt_log} the best-fit spectral parameters and the total $\chi^2$ test statistics. By inspecting the residuals from the fit (see Fig.~\ref{fig:u1700_xrt_bat_spec_t1}), we found no evidence for the presence of a black-body component. Due to the lack of energy coverage in the range 10-15~keV and the limited statistics of the BAT data around 30~keV, we could not evaluate the presence of the possible absorption features revealed by the fit to the \nustar\ data (see earlier in this section). We note, however, that the addition of two Gaussian absorption components with parameters fixed to those determined by \nustar\ did not cause a worsening of the fit and could thus be compatible with the XRT+BAT data.

The second outburst of the source occurred on 2014 December 13 at 01:56:11.09  UT
(image trigger 621278, T$_0=$MJD 57004.08068392; \citealt[][]{kennea14}) and caused
an immediate slew, so that the NFIs were on target at $T_0+$120\,s.
The properties of the \swift{} data of the first observation (ObsID 00621278000)
are strikingly similar to those observed in the first trigger, with a
remarkable X-ray variability and dynamic range (minimum to maximum $\sim 5$--$10$\,c\,s$^{-1}$)
associated with a substantially constant hardness ratio.
Similarly to what was done for the first trigger, we only consider for our
analysis the quasi-simultaneous BAT first orbit spectrum  (T$_0-239$\,s to T$_0+693$\,s)
and the XRT/WT first orbit spectrum (T$_0+120$--$1561$\,s).
The spectra were fit as was done above for the first trigger and we found compatible results (see Table~\ref{u1700:tab:swift_xrt_log}). The same conclusion concerning the presence of the black-body and the Gabs absorption components applies.

The third outburst of the source occurred a few days later on 2015 January 4 at 15:39:06.99  UT
(image trigger 623434, T$_0=$MJD 57026.65216427; \citealt[][]{burrows15}) and caused an immediate slew, so that the NFIs were on target at $T_0+$406\,s.
The \swift{} data of the first observation (ObsID 00623434000) also show the X-ray variability and dynamic range (minimum to maximum $\sim 5$--$130$\,c\,s$^{-1}$) associated by a substantially constant hardness ratio.
We only consider for our analysis the quasi-simultaneous
BAT bright spectrum  (T$_0+0$--$320$\,s)
and the XRT/WT first orbit spectrum (T$_0+406$--$1818$\,s).
The spectra were fit as was done above for the previous triggers, obtaining similar results and the same conclusions concerning additional spectral components beside the absorbed cut-off powerlaw (see Table~\ref{u1700:tab:swift_xrt_log}). As both the combined XRT+BAT spectra of the second and third trigger were relatively similar to those of the first trigger, we did not show for brevity the corresponding images of the best fit models and residuals from the fits.

%
%
%
%
%
\begin{table*} 	
 \tabcolsep 4pt 
\begin{center} 	
 \caption{ 
{\it Swift}/XRT observation log of \srcc: observing sequence, date (MJD of the start of the observation), start and end times (UTC), 
 XRT exposure time, and time since each BAT trigger for the start of the observation.  
We also indicated for each BAT+XRT spectral pair the value of the intercalibration constant, the absorption column density $N_{\rm H}$, the powerlaw photon index $\Gamma$, and the flux in the 1--10~keV energy band (not corrected  for absorption) determined from the spectral fit, as well as  the values of the cutoff energy ($E_\mathrm{C}$) and efolding energy ($E_\mathrm{F}$) of the \texttt(highecut) model. 
}	
 \label{u1700:tab:swift_xrt_log} 	
\scriptsize
 \begin{tabular}{ lllll l  lllllll } 
 \hline 
 \hline 
 \noalign{\smallskip} 
  Sequence   & MJD  & Start time  (UT)                 & End time   (UT)        & Exposure  & Time since & C$_2$ & $N_{\rm H}$    & $\Gamma$  & $F_{\rm 1-10~keV}$                       & $E_\mathrm{C}$ & $E_\mathrm{F}$   & $\chi^2$/d.o.f. \\  
                    &          &(yyyy-mm-dd hh:mm:ss)  & (yyyy-mm-dd hh:mm:ss)  &(s)   & trigger (s)                          &            & $10^{22}$~cm$^{-2}$ &  & 10$^{-11}$ & keV                   & keV  &              \\
 \noalign{\smallskip} 
 \hline 
 \noalign{\smallskip} 
 %
00612974000 BAT&	56917.56128   & 2014-09-17 13:28:15 &      2014-09-17 13:46:48     &     1113  & $-$150  & 
\multirow{2}{*}{$2.64_{-0.76}^{+0.84}$} & 
\multirow{2}{*}{$0.37_{-0.14}^{+0.12}$} &
\multirow{2}{*}{$0.93_{-0.19}^{+0.12}$} &
\multirow{2}{*}{$79.6_{-0.08}^{+0.08}$} & 
\multirow{2}{*}{$4.5_{-2.1}^{+1.4}$}   &
\multirow{2}{*}{$19.6_{-4.94}^{+7.5}$} &
 \multirow{2}{*}{$208.72/203$}  \\
00612974000 WT &	56917.57672	& 2014-09-17 13:50:28	&	2014-09-17 16:38:20	&	629	 &  1183   & & & & & &  \\
\noalign{\smallskip}
%
00621278000 BAT	&     57004.07803   &      2014-12-13 01:52:22     &     2014-12-13 05:37:37     &      13516 &  $-$239 &
\multirow{2}{*}{$1.40_{-0.52}^{+0.81}$} &
\multirow{2}{*}{$0.30_{-0.07}^{+0.07}$} &
\multirow{2}{*}{$0.79_{-0.09}^{+0.08}$} &
\multirow{2}{*}{$108.8_{-3.7}^{+4.5}$} &
\multirow{2}{*}{$5.0_{-0.5}^{+0.6}$} &
\multirow{2}{*}{$7.0_{-1.0}^{+1.3}$} &
\multirow{2}{*}{$501.32/471$} \\
00621278000 WT	&	57004.08219	&	2014-12-13 01:58:20	&	2014-12-13 06:38:33	&	2002 & 120  & & & & & & 	\\
 \noalign{\smallskip}
00623434000 BAT      &     57026.64951    &     2015-01-04 15:35:18	&     2015-01-04 19:12:02     &     13195 & $-$239 & 
\multirow{2}{*}{$1.27_{-0.33}^{+0.43}$} & 
\multirow{2}{*}{$0.23_{-0.06}^{+0.06}$} & 
\multirow{2}{*}{$0.71_{-0.05}^{+0.05}$} & 
\multirow{2}{*}{$210.6_{-9.2}^{+11.2}$} & 
\multirow{2}{*}{$3.87_{-0.52}^{+0.57}$} & 
\multirow{2}{*}{$11.3_{-2.0}^{+2.8}$} & 
\multirow{2}{*}{$574.87/586$} \\
00623434000 WT	&	57026.65698	&	2015-01-04 15:46:02	&	2015-01-04 19:01:06	&	4656 & 406  & & & & & & 	\\
\noalign{\smallskip}
  \hline
  \end{tabular}
  \end{center}
 \end{table*}

\section{\srcb}
\label{sec:j1739}

\begin{figure*}
\centering
\includegraphics[width=12.0cm,angle=-90]{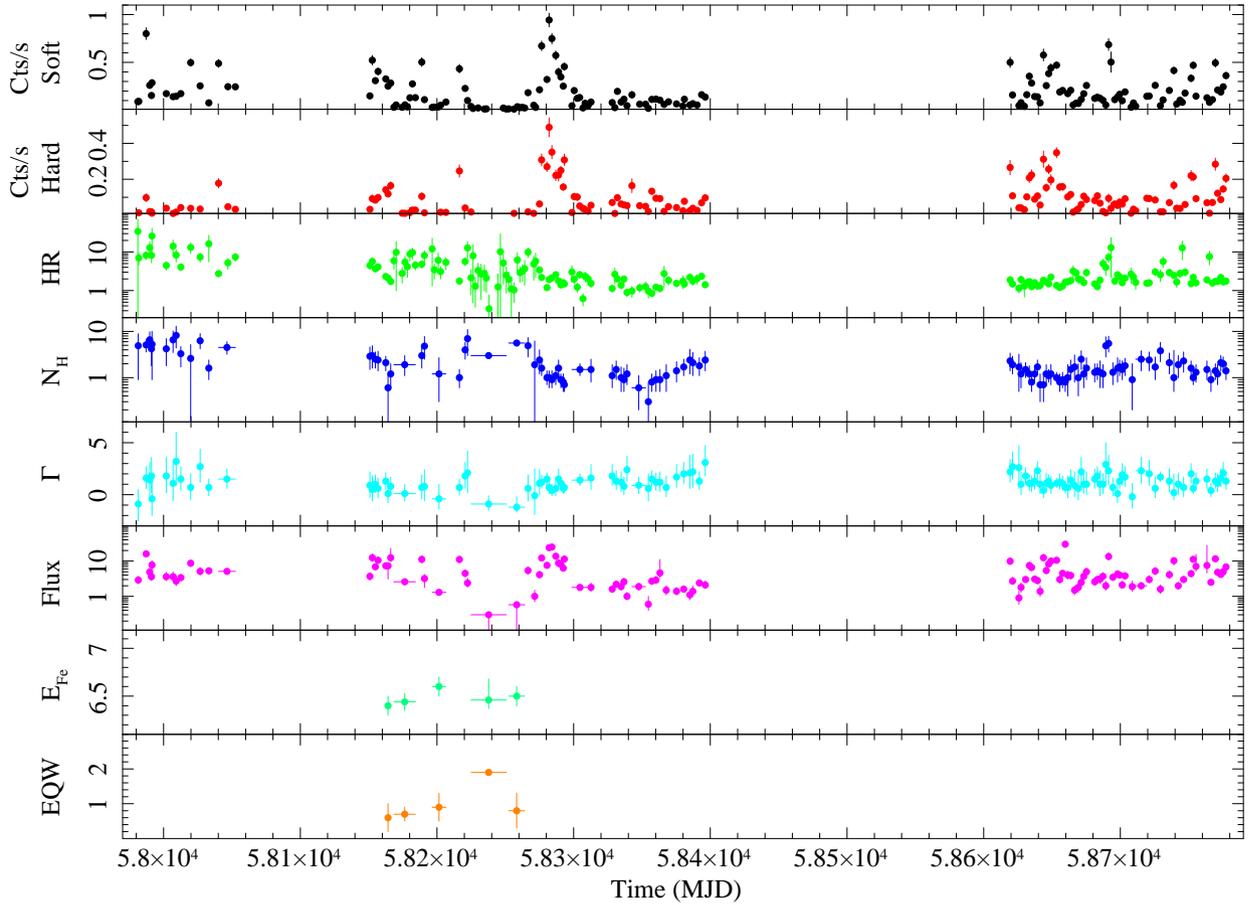}
    \caption{Plot of the energy resolved count-rate (soft is in 0.5-4~keV, while hard is 4-10~keV), HR, absorption column density, power-law photon index, flux, iron line centroid energy and equivalent width as a function of time obtained from the fut to the XRT data of \srcb.\
    Each point in the plot corresponds to one observation in the Table~\ref{i17329:tab:swift_xrt_log}, unless observations have been stacked together as in the table to obtain a sufficient signal-to-noise ratio.}
    \label{fig:17329_xrt_lc}
\end{figure*}

\srcb\ was discovered by \inte\ in 2017 \citep{postel17} and rapidly pointed at with a number of different facilities giving rise to a prompt multi-wavelength campaign. Observations carried out with \xmm\ and \nustar\ revealed the presence of prominent pulsations at  a period of 6680\,s and of a CRSF with a centroid energy of $\sim$20\,keV. These findings confirmed the presence of a NS accretor in this system endowed with a magnetic field of 2.4$\times10^{12}(1+z)$\,G, with $z$ gravitation redshift at the NS surface. The positional accuracy provided by \swift/XRT allowed also for the identification of the optical counterpart through SOAR observations as a late M giant at a distance of 2.7$^{+3.4}_{-1.2}$~kpc. \srcb\ was thus classified as a new member of the SyXBs \citep{bozzo18,bahramian17}. Although no pointed observation with sensitive X-ray telescopes was available before the discovery in 2017, \citet{bozzo18} used the entire available \inte\ archive to show that the source was likely never detectable before in the high energy domain and thus the observations carried out in 2017 might have as well discovered the very first moment when the source shone as a SyXB.

Following the discovery of the source, a long-term monitoring campaign was initiated with \swift\,/XRT in order to investigate the evolution of this object in the X-ray domain. In this paper, we report on the results of this concluded campaign.
\begin{figure}
\centering
\includegraphics[width=6.5cm,angle=-90]{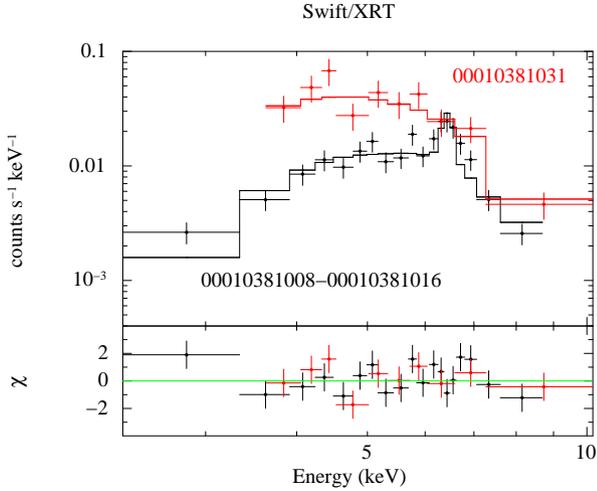}
    \caption{Unfolded spectra of \srcb\ extracted from the XRT observation 00010381031 (red), during which the source displayed a flux close to the average value, and from the combined observations 00010381008-00010381016 (black), during which the source was in one of the faint states that characterize the intra-flare emission around 58200~MJD (see Table~\ref{i17329:tab:swift_xrt_log}). Both spectra could be described by an absorbed powerlaw, but the spectrum in the faint state clearly features a prominent iron line with a centroid energy of $\sim$6.4~keV (see text for details).}
    \label{fig:17329_iron}
\end{figure}
\begin{figure}
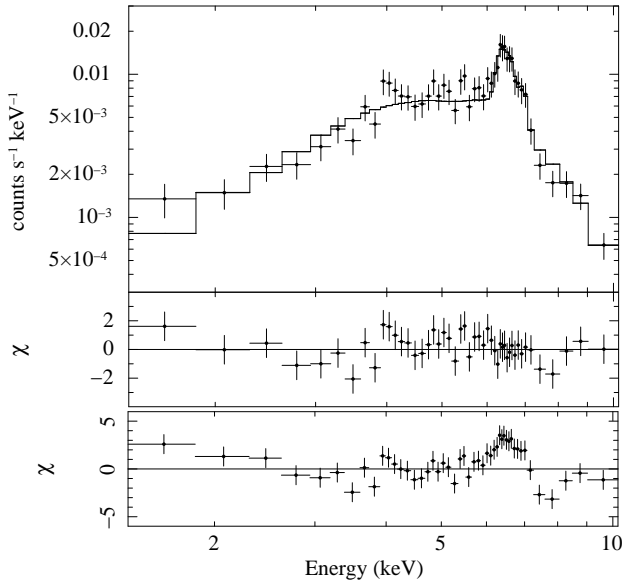

\centering
\includegraphics[width=5.3cm,angle=-90]{17329_xrt_obscure.ps}
\hspace*{0.47cm}\includegraphics[width=2.23cm,angle=-90]{17329_xrt_obscure_residuals.ps}
    \caption{Merged XRT spectrum of \srcb{} obtained by combining together all observations where a iron line could be significantly detected (see Fig.~\ref{fig:17329_xrt_lc} and Table~\ref{i17329:tab:swift_xrt_log}). The best fit model is obtained with an absorbed power-law model and two  Gaussian features at roughly 6.4~keV and 6.7~keV. The middle panel shows the residuals from the best fit when also the additional Gaussian line at roughly 6.9~keV is included, while the bottom panel shows the residuals from the best fit when all Gaussian components are removed.}
    \label{fig:17329_xrt_obscuration}
\end{figure}
The \swift\ data of \srcb\ were uniformly processed and analysed using the standard software ({\sc FTOOLS}\footnote{\href{https://heasarc.gsfc.nasa.gov/ftools/ftools_menu.html}{https://heasarc.gsfc.nasa.gov/ftools/ftools\_menu.html.}} v6.29b), calibration (CALDB\footnote{\href{https://heasarc.gsfc.nasa.gov/docs/heasarc/caldb/caldb_intro.html}{https://heasarc.gsfc.nasa.gov/docs/heasarc/caldb/caldb\_intro.html.}}  20210915), and methods. The \swift/XRT data were filtered with the task {\sc xrtpipeline} (v0.13.6). A log of all available XRT observations of \srcb{} is reported in Table~\ref{i17329:tab:swift_xrt_log}.
%
%
%
%
%
  \setcounter{table}{0}
 \begin{table*}
\tabcolsep 4pt
 \begin{center}
 \caption{{\it Swift}/XRT observation log for IGR~J17329$-$2731. The OBSIDs starting with 0001024 are those already reported by \citet{bozzo18}. In the table we also indicated for each observation the value of the absorption column density $N_{\rm H}$, the powerlaw photon index $\Gamma$, and the flux in the 0.5-10~keV energy band (not corrected  for absorption) determined from the spectral fit. We also reported, where needed, the values of the centroid energy ($E_{\rm Fe}$) and equivalent width (EQW) of the neutron iron line energy. When values of the fit cover more than one row, it means that the corresponding observations have been merged together due to the otherwise too low number of counts to perform the spectral extraction.  
\label{i17329:tab:swift_xrt_log}
}
\scriptsize
\begin{tabular}{ lllllllllll }
 \hline
 \hline
 \noalign{\smallskip}
 Sequence        & MJD                      & Start time  (UT)                     & End time   (UT)                      & Exposure  & $N_{\rm H}$ & $\Gamma$ & $F_{\rm 0.5-10~keV}$ & $E_{\rm Fe}$ & EQW & Cstat/d.o.f. \\
                        &                             & (yyyy-mm-dd hh:mm:ss)     & (yyyy-mm-dd hh:mm:ss)      &(s)  & 10$^{23}$~cm$^{-2}$ & & 10$^{-11}$~erg~cm$^{-2}$~s$^{-1}$ & keV & keV  & ($\chi^2$/d.o.f.)          \\
  \noalign{\smallskip}
 \hline
 \noalign{\smallskip}
00010244001	&	57981.10361	&	2017-08-16 02:29:12	&	2017-08-16 02:45:47	&	\multirow{2}{*}{1973}   &       \multirow{2}{*}{4.9$\pm$4.0}    &       \multirow{2}{*}{-0.9$^{+1.4}_{-1.6}$}  &      \multirow{2}{*}{2.9$^{+0.8}_{-0.6}$}    & & &  \multirow{2}{*}{39.6/23} \\
00010244002	&	57981.71182	&	2017-08-16 17:05:01	&	2017-08-16 20:02:18	&	& & & & & & \\
00010244003	&	57987.15883	&	2017-08-22 03:48:43	&	2017-08-22 05:29:59	&	985	&       5.1$^{+1.7}_{-1.5}$     &       1.6$^{+1.1}_{-1.0}$         &       16.0$\pm$3.0            & & & 55.8/49 \\
00010244004	&	57989.69705	&	2017-08-24 16:43:44	&	2017-08-24 23:10:53	&	893	&       6.6$^{+3.2}_{-2.7}$     &       1.4$^{+1.7}_{-1.5}$         &       4.9$^{+1.4}_{-0.9}$     &  & & 28.8/26 \\
00010244005	&	57990.95983	&	2017-08-25 23:02:08	&	2017-08-26 06:38:53	&	602	 &       4.2$^{+2.8}_{-2.4}$     &       1.8$^{+1.8}_{-1.6}$         &       3.6$^{+1.2}_{-0.8}$     &  & & 25.9/18  \\
00010244006	&	57991.33843	&	2017-08-26 08:07:20	&	2017-08-26 14:58:53	&	865	 & 5.3$^{+4.7}_{-4.4}$     &       -0.4$^{+1.7}_{-1.6}$         &       7.8$^{+2.5}_{-2.0}$     &  & & 13.7/15 \\
00010244007	&	58001.89720	&	2017-09-05 21:31:57	&	2017-09-05 23:16:53	&	659 &       4.2$^{+2.8}_{-2.4}$     &       1.8$^{+1.8}_{-1.6}$         &       3.6$^{+1.2}_{-0.8}$     &  & & 25.9/18      \\
00010244008	&	58006.94238	&	2017-09-10 22:37:01	&	2017-09-10 22:52:54	&	953	 &       6.6$^{+3.6}_{-3.1}$     &       1.1$^{+1.9}_{-1.7}$         &       3.6$^{+1.3}_{-0.9}$     &  & & 18.0/19 \\
00010244009	&	58009.06811	&	2017-09-13 01:38:04	&	2017-09-13 06:27:52	&	717	&       8.2$^{+4.8}_{-3.8}$     &       3.2$^{+2.8}_{-2.3}$         &       2.7$^{+1.2}_{-0.7}$     &  & & 19.7/14 \\
00010244010	&	58012.53078	&	2017-09-16 12:44:19	&	2017-09-16 13:01:52	&	1053 &       3.3$^{+1.4}_{-1.6}$     &       1.5$^{+1.2}_{-1.1}$         &       3.4$^{+0.8}_{-0.6}$     &  & & 51.8/29 		\\
00010244011	&	58019.63368	&	2017-09-23 15:12:29	&	2017-09-23 21:47:54	&	898	&       2.6$^{+2.6}_{-2.5}$     &       0.7$^{+1.3}_{-1.2}$         &       8.7$^{+1.9}_{-1.7}$     &  & & 23.3/32\\
00010244012	&	58026.68470	&	2017-09-30 16:25:57	&	2017-09-30 16:41:53	&	955	&       6.3$^{+2.7}_{-2.1}$     &       2.7$^{+1.7}_{-1.5}$         &       5.1$^{+1.5}_{-1.0}$     &  & & 32.2/36 \\
00010244013	&	58033.05169	&	2017-10-07 01:14:26	&	2017-10-07 02:56:53	&	\multirow{2}{*}{1508}   &       \multirow{2}{*}{1.6$^{+0.8}_{-0.7}$}    &       \multirow{2}{*}{0.7$\pm$0.8}  &      \multirow{2}{*}{5.3$^{+1.1}_{-0.9}$}    & & &  \multirow{2}{*}{72.8/44} \\
00010244014	&	58040.16153	&	2017-10-14 03:52:36	&	2017-10-14 04:02:53	&		& & & &  & & \\
00010244015	&	58047.01855	&	2017-10-21 00:26:42	&	2017-10-21 01:55:44	& \multirow{2}{*}{1381}   &       \multirow{2}{*}{4.5$^{+1.5}_{-1.3}$}    &       \multirow{2}{*}{1.5$^{+1.0}_{-0.9}$}  &      \multirow{2}{*}{5.1$\pm$0.8}    & & &  \multirow{2}{*}{53.7/57} \\
00010244016	&	58052.52036	&	2017-10-26 12:29:18	&	2017-10-26 12:44:54	&	& & & &  & & \\
%
00010381001	&	58150.88910	&	2018-02-01 21:20:17	&	2018-02-01 21:36:53	&	995	&       2.9$^{+1.7}_{-1.4}$     &       0.9$^{+1.3}_{-1.2}$         &       3.7$^{+1.0}_{-0.8}$      & & & 104.7/131 \\
00010381002	&	58152.68608	&	2018-02-03 16:27:56	&	2018-02-03 16:43:52	&	955	&       3.0$^{+1.9}_{-1.4}$     &       0.5$^{+1.3}_{-1.1}$         &       12.5$^{+3.4}_{-2.7}$      & & & 95.0/115 \\
00010381003	&	58154.94023	&	2018-02-05 22:33:55	&	2018-02-05 22:51:53	&	1078 &       2.5$^{+1.1}_{-0.9}$     &       1.0$^{+0.9}_{-0.8}$         &       6.9$^{+1.4}_{-1.2}$      & & & 107.2/183 	\\
00010381004	&	58156.87057	&	2018-02-07 20:53:36	&	2018-02-07 21:08:52	&	915	&       2.4$^{+1.3}_{-1.0}$     &       0.6$^{+1.0}_{-0.9}$         &       10.6$^{+2.4}_{-2.0}$      & & & 124.5/158\\
00010381005	&	58162.51007	&	2018-02-13 12:14:30	&	2018-02-13 12:30:53	&	983	&       2.1$^{+0.8}_{-0.7}$     &       1.3$^{+0.8}_{-0.8}$         &       7.4$^{+1.4}_{-1.1}$      & & & 180.2/209\\
00010381006	&	58164.16929	&	2018-02-15 04:03:46	&	2018-02-15 04:19:51	&	965	&       0.6$^{+0.5}_{-0.6}$     &       0.1$^{+1.1}_{-0.9}$         &       7.3$^{+0.3}_{-4.2}$      & 6.4$^{+0.1}_{-0.1}$ & 0.6$^{-0.4}_{+0.4}$ & (6.9/11)\\ 
00010381007	&	58166.16069	&	2018-02-17 03:51:23	&	2018-02-17 04:07:53	&	990	&       1.2$^{+0.8}_{-0.6}$     &       0.8$^{+0.9}_{-0.8}$         &       12.5$^{+10.5}_{-2.7}$     & &  & 123.0/148 \\ 
00010381008	&	58168.48439	&	2018-02-19 11:37:31	&	2018-02-19 12:13:39	&	\multirow{8}{*}{8834}   &       \multirow{8}{*}{1.9$^{+1.1}_{-0.8}$}    &       \multirow{8}{*}{0.1$^{+0.9}_{-0.7}$}  &      \multirow{8}{*}{2.6$\pm$0.2}    & \multirow{8}{*}{6.44$^{+0.09}_{-0.09}$} & \multirow{8}{*}{0.7$^{+0.2}_{-0.2}$} &  \multirow{8}{*}{(20.7/19)} \\     
00010381009	&	58170.14349	&	2018-02-21 03:26:37	&	2018-02-21 03:42:52	&		& & & &  & & \\
00010381011	&	58174.52953	&	2018-02-25 12:42:31	&	2018-02-25 12:58:54	&		& & & &  & & \\
00010381012	&	58176.58920	&	2018-02-27 14:08:26	&	2018-02-27 14:23:54	&		& & & &  & & \\
00010381013	&	58178.31495	&	2018-03-01 07:33:31	&	2018-03-01 07:48:54	&		& & & &  & & \\
00010381014	&	58180.24239	&	2018-03-03 05:49:02	&	2018-03-03 06:03:52	&		& & & &  & & \\
00010381015	&	58182.03505	&	2018-03-05 00:50:27	&	2018-03-05 01:07:53	& 	    & & & &  & & \\
00010381016	&	58184.22645	&	2018-03-07 05:26:05	&	2018-03-07 05:41:52	&		& & & &  & &  \\
00010381017	&	58188.87579	&	2018-03-11 21:01:08	&	2018-03-11 21:17:53	&	1005 &       3.0$^{+1.8}_{-1.4}$     &       0.7$^{+1.2}_{-1.1}$         &       11.2$^{+2.8}_{-2.3}$     & &  & 134.3/132\\
00010381018	&	58190.80228	&	2018-03-13 19:15:16	&	2018-03-13 19:33:52	&	1116 &       4.8$^{+3.0}_{-2.3}$     &       0.8$^{+1.6}_{-1.4}$         &       3.2$^{+1.0}_{-1.4}$     & &  & 72.42/95	\\ 
00010381020	&	58196.51638	&	2018-03-19 12:23:35	&	2018-03-19 12:38:52	&	\multirow{5}{*}{4705}   &       \multirow{5}{*}{1.2$^{+1.5}_{-0.9}$}    &       \multirow{5}{*}{-0.4$^{+1.4}_{-1.0}$}  &      \multirow{5}{*}{1.3$\pm$0.1}    & \multirow{5}{*}{6.6$^{+0.1}_{-0.1}$} & \multirow{5}{*}{0.9$^{+0.4}_{-0.4}$} &  \multirow{5}{*}{(4.8/7)} \\  
00010381021	&	58198.03978	&	2018-03-21 00:57:17	&	2018-03-21 01:14:52	&		& & & &  & &  	\\
00010381022	&	58200.90118	&	2018-03-23 21:37:42	&	2018-03-23 21:54:52	&		& & & &  & &  	\\
00010381023	&	58202.63502	&	2018-03-25 15:14:25	&	2018-03-25 15:28:53	&		& & & &  & &  	\\
00010381025	&	58206.54025	&	2018-03-29 12:57:57	&	2018-03-29 13:11:52	&		& & & &  & &  \\
00010381029	&	58216.36885	&	2018-04-08 08:51:08	&	2018-04-08 09:09:53	&	1126 &       1.0$^{+0.5}_{-0.4}$     &       0.7$^{+0.8}_{-0.7}$         &       11.1$^{+2.6}_{-2.1}$     &  & & 114.1/141	\\
00010381031	&	58220.62044	&	2018-04-12 14:53:25	&	2018-04-12 15:11:54	&	1108 &       4.0$^{+1.8}_{-1.5}$     &       1.8$^{+1.3}_{-1.2}$         &       4.5$^{+1.1}_{-0.9}$     &  & & 99.4/143	\\
00010381032	&	58222.34786	&	2018-04-14 08:20:54	&	2018-04-14 08:36:52	&	958	&       7.0$^{+4.0}_{-4.0}$     &       2.1$^{+2.1}_{-1.9}$         &       2.4$^{+0.9}_{-0.6}$     &  & & 57.7/80	\\\\
00010381033	&	58224.87328	&	2018-04-16 20:57:31	&	2018-04-16 21:01:52	&	\multirow{12}{*}{10659}   &       \multirow{12}{*}{$<3.0$}    &       \multirow{12}{*}{-0.9$^{+0.8}_{-0.6}$}  &      \multirow{12}{*}{0.30$^{+0.05}_{-0.25}$}    & \multirow{12}{*}{6.46$^{+0.22}_{-0.09}$} & \multirow{12}{*}{1.9$^{+0.07}_{-0.08}$} &  \multirow{12}{*}{19.8/26} \\  
00010381034	&	58226.40209	&	2018-04-18 09:39:00	&	2018-04-18 09:54:53	&		& & & &  & &  	\\
00010381035	&	58228.00227	&	2018-04-20 00:03:15	&	2018-04-20 00:17:52	&		& & & &  & &  	\\
00010381036	&	58230.12174	&	2018-04-22 02:55:17	&	2018-04-22 03:10:53	&			& & & &  & &  	\\
00010381037	&	58232.24687	&	2018-04-24 05:55:29	&	2018-04-24 06:10:52	&		& & & &  & &  	\\
00010381038	&	58234.63747	&	2018-04-26 15:17:57	&	2018-04-26 15:34:52	&		& & & &  & &  		\\
00010381039	&	58236.10438	&	2018-04-28 02:30:18	&	2018-04-28 02:45:53	&		& & & &  & &  	\\
00010381040	&	58238.10514	&	2018-04-30 02:31:24	&	2018-04-30 02:46:52	&		& & & &  & &  	\\
00010381041	&	58244.66515	&	2018-05-06 15:57:49	&	2018-05-06 16:13:51	&		& & & &  & &  	\\
00010381042	&	58246.39222	&	2018-05-08 09:24:47	&	2018-05-08 11:08:53	&		& & & &  & &  	\\
00010381043	&	58248.51643	&	2018-05-10 12:23:39	&	2018-05-10 12:39:52	& 		& & & &  & &   \\
00010381044	&	58250.64859	&	2018-05-12 15:33:57	&	2018-05-12 15:49:53	&		& & & &  & &  	\\
 \noalign{\smallskip}
  \hline
  \end{tabular}
  \end{center}
  \end{table*}

\setcounter{table}{0}
 \begin{table*}
\tabcolsep 4pt
 \begin{center}
 \caption{Continued.}
\scriptsize
\begin{tabular}{ lllllllllll }
 \hline
 \hline
 Sequence        & MJD                      & Start time  (UT)                     & End time   (UT)                      & Exposure  & $N_{\rm H}$ & $\Gamma$ & $F_{\rm 0.5-10~keV}$ & $E_{\rm Fe}$ & EQW & Cstat/d.o.f.\\
                        &                             & (yyyy-mm-dd hh:mm:ss)     & (yyyy-mm-dd hh:mm:ss)      &(s)  & 10$^{23}$~cm$^{-2}$ & & 10$^{-11}$~erg~cm$^{-2}$~s$^{-1}$ & keV & keV           \\
  \noalign{\smallskip}
 \hline
 \noalign{\smallskip}
00010381045	&	58252.50513	&	2018-05-14 12:07:23	&	2018-05-14 12:24:53	&	\multirow{7}{*}{6907}   &       \multirow{7}{*}{$<5.6$}    &       \multirow{7}{*}{-1.2$^{+1.0}_{-0.5}$}  &      \multirow{7}{*}{0.58$^{+0.05}_{-0.54}$}    & \multirow{7}{*}{6.5$^{+0.1}_{-0.1}$} & \multirow{7}{*}{0.8$^{+0.5}_{-0.5}$} &  \multirow{7}{*}{18.6/27}	\\
00010381046	&	58254.43585	&	2018-05-16 10:27:37	&	2018-05-16 10:43:52	&		& & & &  & & 		\\
00010381047	&	58256.36750	&	2018-05-18 08:49:12	&	2018-05-18 09:05:52	&		& & & &  & & 	\\
00010381048	&	58258.82480	&	2018-05-20 19:47:42	&	2018-05-20 20:03:52	&		& & & &  & & 		\\
00010381049	&	58260.74879	&	2018-05-22 17:58:15	&	2018-05-22 18:15:53	&		& & & &  & & 		\\
00010381050	&	58262.46195	&	2018-05-24 11:05:12	&	2018-05-24 11:20:52	&		& & & &  & & 	\\
00010381051	&	58264.18867	&	2018-05-26 04:31:41	&	2018-05-26 04:46:54	&		& & & &  & & \\
00010381052	&	58266.58874	&	2018-05-28 14:07:46	&	2018-05-28 14:24:52	&	1025 &       4.9$^{+2.5}_{-2.1}$     &       0.6$^{+1.3}_{-1.2}$         &       5.4$^{+1.4}_{-1.1}$     &  & & 106.5/131	\\
00010381053	&	58270.84194	&	2018-06-01 20:12:23	&	2018-06-01 20:30:54	 & \multirow{2}{*}{2061}   &       \multirow{2}{*}{1.9$^{+4.3}_{-1.8}$}    &       \multirow{2}{*}{-0.1$^{+2.6}_{-1.8}$}  &      \multirow{2}{*}{1.0$^{+0.5}_{-0.3}$}    & & &  \multirow{2}{*}{42.2/42} \\
00010381054	&	58272.16777	&	2018-06-03 04:01:35	&	2018-06-03 04:17:53	&		& & & &  & & 		\\
00010381055	&	58274.96290	&	2018-06-05 23:06:34	&	2018-06-05 23:22:52	&	978	&       2.4$^{+1.6}_{-1.2}$     &       1.1$^{+1.3}_{-1.1}$         &       4.1$^{+1.1}_{-0.8}$     &  & & 110.4/146	\\
00010381056	&	58276.54765	&	2018-06-07 13:08:36	&	2018-06-07 15:03:52	&	890		&       1.6$^{+0.6}_{-0.5}$     &       1.2$^{+0.7}_{-0.6}$         &       12.2$^{+2.1}_{-1.8}$     &  & & 202.4/217\\
00010381057	&	58280.46373	&	2018-06-11 11:07:46	&	2018-06-11 11:25:54	&	1088 		&       1.0$^{+0.4}_{-0.3}$     &       1.5$^{+0.6}_{-0.6}$         &       7.6$^{+1.4}_{-1.1}$     &  & & 192.4/191	\\
00010381058	&	58282.05881	&	2018-06-13 01:24:40	&	2018-06-13 01:40:53	&	973			&       1.0$^{+0.4}_{-0.4}$     &       0.7$^{+0.6}_{-0.6}$         &       24.0$^{+4.0}_{-4.0}$     &  & & 176.0/225\\
00010381059	&	58284.05580	&	2018-06-15 01:20:20	&	2018-06-15 01:36:53	&	993			&       0.9$^{+0.5}_{-0.3}$     &       0.4$^{+0.6}_{-0.5}$         &       25.0$^{+5.0}_{-3.0}$     &  & & 173.8/251\\
00010381060	&	58286.90682	&	2018-06-17 21:45:49	&	2018-06-17 22:00:52	&	903			&       1.1$^{+0.6}_{-0.4}$     &       0.6$^{+0.8}_{-0.7}$         &       13.7$^{+2.8}_{-2.3}$     &  & & 148.5/176\\
00010381061	&	58288.97411	&	2018-06-19 23:22:43	&	2018-06-19 23:37:53	&	910			&       1.6$^{+0.8}_{-0.6}$     &       1.5$^{+1.0}_{-0.9}$         &       8.8$^{+2.3}_{-1.8}$     &  & & 110.8/118\\
00010381062	&	58290.37278	&	2018-06-21 08:56:35	&	2018-06-21 18:43:52	&	1171 		&       0.9$^{+0.3}_{-0.3}$     &       1.1$^{+0.6}_{-0.5}$         &       8.1$^{+1.4}_{-1.1}$     &  & & 179.6/237	\\
00010381063	&	58292.35199	&	2018-06-23 08:26:52	&	2018-06-23 08:42:52	&	960			&       0.8$^{+0.4}_{-0.3}$     &       0.6$^{+0.6}_{-0.5}$         &       6.3$^{+1.2}_{-1.0}$     &  & & 147.1/197\\
00010381064	&	58293.14907	&	2018-06-24 03:34:39	&	2018-06-24 03:45:53	&	674			&       0.7$^{+0.3}_{-0.2}$     &       0.7$^{+0.5}_{-0.5}$         &       11.4$^{+2.0}_{-1.7}$     &  & & 162.1/219\\
00010381065	&	58298.73917	&	2018-06-29 17:44:24	&	2018-06-29 17:53:53	&	\multirow{7}{*}{4679}   &       \multirow{7}{*}{1.5$^{+0.5}_{-0.4}$}    &       \multirow{7}{*}{1.4$^{+0.6}_{-0.5}$}  &      \multirow{7}{*}{1.8$^{+0.3}_{-0.2}$}    & & &  \multirow{7}{*}{239.5/285} \\
00010381066	&	58300.40880	&	2018-07-01 09:48:40	&	2018-07-01 11:14:53	&		& & & &  & & 		\\	
00010381067	&	58302.58538	&	2018-07-03 14:02:56	&	2018-07-03 15:38:52	&			& & & &  & & 		\\
00010381068	&	58304.32606	&	2018-07-05 07:49:31	&	2018-07-05 09:25:54	&			& & & &  & & 		\\
00010381069	&	58306.78253	&	2018-07-07 18:46:50	&	2018-07-07 20:13:54	&		& & & &  & & 		\\
00010381070	&	58308.36775	&	2018-07-09 08:49:33	&	2018-07-09 09:01:52	&		& & & &  & & 		\\
00010381071	&	58310.49337	&	2018-07-11 11:50:26	&	2018-07-11 12:06:52	&		& & & &  & & 		\\	
00010381072	&	58312.68326	&	2018-07-13 16:23:53	&	2018-07-13 16:41:54	&	1081 		&       1.5$^{+1.0}_{-0.7}$     &       1.6$^{+1.3}_{-1.1}$         &       1.8$^{+0.6}_{-0.4}$     &  & & 80.9/94	\\
00010381074	&	58328.21535	&	2018-07-29 05:10:06	&	2018-07-29 05:28:52	&	1126 		&       1.1$^{+0.8}_{-0.5}$     &       1.8$^{+1.1}_{-0.9}$         &       1.6$^{+0.4}_{-0.3}$     &  & & 73.8/110	\\
00010381075	&	58330.14540	&	2018-07-31 03:29:22	&	2018-07-31 05:12:53	&	\multirow{2}{*}{1983}   &       \multirow{2}{*}{1.5$^{+0.8}_{-0.6}$}    &       \multirow{2}{*}{1.3$^{+1.0}_{-0.8}$}  &      \multirow{2}{*}{2.2$^{+0.5}_{-0.4}$}    & & &  \multirow{2}{*}{130.5/148} \\	
00010381076	&	58332.01371	&	2018-08-02 00:19:44	&	2018-08-02 00:35:52	&	& & & &  & & 		\\
00010381077	&	58334.73540	&	2018-08-04 17:38:58	&	2018-08-04 17:56:54	&	1076	 		&       1.0$^{+0.7}_{-0.5}$     &       1.2$^{+1.0}_{-0.9}$         &       1.9$^{+0.6}_{-0.4}$     &  & & 87.5/101	\\
00010381078	&	58336.79321	&	2018-08-06 19:02:13	&	2018-08-06 19:19:54	&	1061  		&       0.9$^{+0.7}_{-0.4}$     &       0.8$^{+0.9}_{-0.8}$         &       2.6$^{+0.7}_{-0.5}$     &  & & 98.1/124	\\
%
%
00010381079	&	58339.04198	&	2018-08-09 01:00:27	&	2018-08-09 01:16:52	&	985	 		&       1.2$^{+0.7}_{-0.5}$     &       2.4$^{+1.3}_{-1.1}$         &       1.0$^{+0.3}_{-0.2}$     &  & & 73.7/72	\\
00010381080	&	58342.69665	&	2018-08-12 16:43:10	&	2018-08-12 16:46:53	&	\multirow{3}{*}{1682}   &       \multirow{3}{*}{0.6$^{+0.5}_{-0.4}$}    &       \multirow{3}{*}{0.9$^{+0.8}_{-0.8}$}  &      \multirow{3}{*}{1.9$^{+0.5}_{-0.4}$}    & & &  \multirow{3}{*}{126.2/139} \\
00010381081	&	58348.28656	&	2018-08-18 06:52:39	&	2018-08-18 07:07:54	&	& & & &  & & 		\\
00010381082	&	58352.59571	&	2018-08-22 14:17:49	&	2018-08-22 14:26:53	&	& & & &  & & 		\\
00010381083	&	58354.39100	&	2018-08-24 09:23:02	&	2018-08-24 22:29:52	&	1066 	 		&       0.3$^{+0.5}_{-0.2}$     &       0.6$^{+1.4}_{-1.2}$         &       0.6$^{+0.4}_{-0.2}$     &  & & 49.9/33	\\
00010381084	&	58357.17574	&	2018-08-27 04:13:03	&	2018-08-27 05:56:53	&	1038	 		&       0.8$^{+0.4}_{-0.3}$     &       1.5$^{+0.8}_{-0.7}$         &       2.7$^{+0.6}_{-0.5}$     &  & & 125.9/151	\\
00010381085	&	58360.23100	&	2018-08-30 05:32:38	&	2018-08-30 05:49:53	&	1036 	 		&       0.9$^{+0.5}_{-0.4}$     &       1.2$^{+0.9}_{-0.8}$         &       2.9$^{+0.8}_{-0.6}$     &  & & 120.3/146	\\
00010381086	&	58363.14928	&	2018-09-02 03:34:57	&	2018-09-02 03:48:52	&	832		 		&       0.9$^{+0.6}_{-0.4}$     &       1.2$^{+1.0}_{-0.9}$         &       4.6$^{+6.5}_{-1.2}$     &  & & 118.0/98	\\
00010381087	&	58366.13522	&	2018-09-05 03:14:43	&	2018-09-05 03:19:16	&	\multirow{2}{*}{1385}   &       \multirow{2}{*}{1.1$^{+0.8}_{-0.6}$}    &       \multirow{2}{*}{0.7$^{+1.1}_{-0.9}$}  &      \multirow{2}{*}{1.5$^{+0.5}_{-0.4}$}    & & &  \multirow{2}{*}{76.7/83} \\	
00010381088	&	58369.39329	&	2018-09-08 09:26:20	&	2018-09-08 09:44:53	& 	& & & &  & & 		\\
00010381089	&	58375.16760	&	2018-09-14 04:01:20	&	2018-09-14 08:55:53	&	1271			 		&       1.4$^{+0.8}_{-0.6}$     &       1.7$^{+1.2}_{-1.0}$         &       1.4$^{+0.4}_{-0.3}$     &  & & 90.3/98	\\
00010381090	&	58379.95453	&	2018-09-18 22:54:30	&	2018-09-18 23:09:53	&	\multirow{2}{*}{1938}   &       \multirow{2}{*}{1.7$^{+0.8}_{-0.6}$}    &       \multirow{2}{*}{2.0$^{+1.1}_{-0.9}$}  &      \multirow{2}{*}{1.6$^{+0.4}_{-0.3}$}    & & &  \multirow{2}{*}{136.3/147} \\
00010381091	&	58381.01428	&	2018-09-20 00:20:33	&	2018-09-20 02:14:52	&	& & & &  & & 		\\
00010381092	&	58384.07019	&	2018-09-23 01:41:04	&	2018-09-24 21:08:52	&	1447	&       2.4$^{+1.7}_{-1.2}$     &       2.1$^{+1.7}_{-1.5}$         &       1.1$^{+0.4}_{-0.3}$     &  & & 81.6/81	\\
00010381093	&	58387.19114	&	2018-09-26 04:35:14	&	2018-09-26 04:52:52	&	1058 &       2.1$^{+1.4}_{-1.0}$     &       2.2$^{+1.7}_{-1.5}$         &       1.4$^{+0.6}_{-0.4}$     &  & & 60.2/79	\\	
00010381094	&	58390.25699	&	2018-09-29 06:10:04	&	2018-09-29 06:25:54	&	\multirow{2}{*}{2166}   &       \multirow{2}{*}{1.8$^{+0.9}_{-0.7}$}    &       \multirow{2}{*}{1.3$^{+0.9}_{-0.8}$}  &      \multirow{2}{*}{2.4$^{+0.5}_{-0.4}$}    & & &  \multirow{2}{*}{165.7/178}  \\ 
00010381095	&	58393.43297	&	2018-10-02 10:23:28	&	2018-10-02 18:29:53	&	& & & &  & & 		\\
00010381096	&	58396.22600	&	2018-10-05 05:25:26	&	2018-10-05 05:41:52	&	985	&       2.4$^{+1.3}_{-1.0}$     &       3.1$^{+1.6}_{-1.3}$         &       2.1$^{+0.6}_{-0.4}$     &  & & 98.3/118	\\
%
%
00010381097	&	58619.35967	&	2019-05-16 08:37:55	&	2019-05-16 08:51:52	&	837	&       2.3$^{+1.1}_{-0.9}$     &       2.2$^{+1.2}_{-1.0}$         &       9.9$^{+2.5}_{-1.9}$     &  & & 124.1/116	\\
00010381098	&	58621.08933	&	2019-05-18 02:08:38	&	2019-05-18 02:23:53	&	915	&       1.9$^{+1.0}_{-0.7}$     &       2.7$^{+1.4}_{-1.2}$         &       2.7$^{+0.8}_{-0.5}$     &  & & 116.3/113	\\
00010381099	&	58625.72455	&	2019-05-22 17:23:21	&	2019-05-22 17:39:54	&	993	&       1.7$^{+1.5}_{-1.2}$     &       2.6$^{+2.1}_{-2.0}$         &       0.9$^{+0.5}_{-0.3}$     &  & & 38.5/58	\\
00010381100	&	58627.31490	&	2019-05-24 07:33:27	&	2019-05-24 07:48:52	&	925	&       1.2$^{+0.9}_{-0.6}$     &       1.0$^{+1.2}_{-1.0}$         &       1.8$^{+0.6}_{-0.5}$     &  & & 72.5/77	\\
00010381101	&	58629.78721	&	2019-05-26 18:50:38	&	2019-05-26 19:06:54	&	\multirow{2}{*}{1429}   &       \multirow{2}{*}{1.5$^{+0.6}_{-0.5}$}    &       \multirow{2}{*}{1.8$^{+0.8}_{-0.8}$}  &      \multirow{2}{*}{3.0$^{+0.6}_{-0.5}$}    & & &  \multirow{2}{*}{106.5/182}  \\
00010381102	&	58631.49904	&	2019-05-28 11:58:37	&	2019-05-28 12:16:53	&		& & & &  & & 		 \\
00010381103	&	58633.35948	&	2019-05-30 08:37:38	&	2019-05-30 08:54:51	&	1033	&       1.2$^{+0.6}_{-0.5}$     &       1.1$^{+0.8}_{-0.7}$         &       7.5$^{+1.7}_{-1.4}$     &  & & 142.8/160	\\
00010381104	&	58635.03407	&	2019-06-01 00:49:03	&	2019-06-01 01:04:54	&	950	&       0.8$^{+0.4}_{-0.3}$     &       1.1$^{+0.7}_{-0.8}$         &       6.6$^{+1.5}_{-1.2}$     &  & & 123.3/153	\\
00010381105	&	58637.27834	&	2019-06-03 06:40:48	&	2019-06-03 06:58:54	&	1086 &       1.2$^{+0.6}_{-0.4}$     &       1.3$^{+0.9}_{-0.8}$         &       3.1$^{+0.7}_{-0.6}$     &  & & 123.3/145	\\
 \noalign{\smallskip}
  \hline
  \end{tabular}
  \end{center}
  \end{table*}

\setcounter{table}{0}
 \begin{table*}
\tabcolsep 4pt
 \begin{center}
 \caption{Continued.}
\scriptsize
\begin{tabular}{ lllllllllll }
 \hline
 \hline
 Sequence        & MJD                      & Start time  (UT)                     & End time   (UT)                      & Exposure  & $N_{\rm H}$ & $\Gamma$ & $F_{\rm 0.5-10~keV}$ & $E_{\rm Fe}$ & EQW & Cstat/d.o.f.\\
                        &                             & (yyyy-mm-dd hh:mm:ss)     & (yyyy-mm-dd hh:mm:ss)      &(s)  & 10$^{23}$~cm$^{-2}$ & & 10$^{-11}$~erg~cm$^{-2}$~s$^{-1}$ & keV & keV           \\
  \noalign{\smallskip}
 \hline
 \noalign{\smallskip}
 00010381106	&	58639.34760	&	2019-06-05 08:20:32	&	2019-06-05 08:38:53	&	1101 &       1.7$^{+0.7}_{-0.5}$     &       2.3$^{+0.9}_{-0.8}$         &       2.8$^{+0.6}_{-0.5}$     &  & & 116.2/149	\\
00010381107	&	58641.34120	&	2019-06-07 08:11:19	&	2019-06-07 08:24:52	&	812	&       0.7$^{+0.5}_{-0.4}$     &       1.0$^{+1.0}_{-0.9}$         &       1.4$^{+0.6}_{-0.4}$     &  & & 74.2/77	\\
00010381108	&	58643.72603	&	2019-06-09 17:25:29	&	2019-06-09 17:43:52	&	1103	&       0.7$^{+0.6}_{-0.4}$     &       0.4$^{+0.8}_{-0.7}$         &       12.5$^{+3.2}_{-2.5}$     &  & & 136.0/119	\\
00010381109	&	58645.78926	&	2019-06-11 18:56:32	&	2019-06-11 19:13:52	&	725	&       1.2$^{+0.7}_{-0.5}$     &       1.1$^{+1.0}_{-0.8}$         &       5.4$^{+1.3}_{-1.0}$     &  & & 114.7/135	\\
00010381110	&	58647.57200	&	2019-06-13 13:43:41	&	2019-06-13 14:01:54	&	1093&       1.1$^{+0.4}_{-0.3}$     &       1.3$^{+0.6}_{-0.6}$         &       8.3$^{+1.4}_{-1.2}$     &  & & 208.7/211		\\
00010381111	&	58649.23274	&	2019-06-15 05:35:08	&	2019-06-15 05:51:54	&	1005	&       1.2$^{+0.7}_{-0.5}$     &       0.9$^{+0.9}_{-0.8}$         &       10.0$^{+2.4}_{-2.0}$     &  & & 130.7/139	\\
00010381112	&	58653.34838	&	2019-06-19 08:21:39	&	2019-06-19 08:37:52	&	973	&       1.0$^{+0.3}_{-0.2}$     &       1.1$^{+0.5}_{-0.4}$         &       10.6$^{+1.4}_{-1.2}$     &  & & 251.5/319	\\
00010381113	&	58655.67261	&	2019-06-21 16:08:33	&	2019-06-21 16:24:54	&	980	&       0.8$^{+0.4}_{-0.3}$     &       1.2$^{+0.8}_{-0.67}$         &       3.0$^{+0.8}_{-0.6}$     &  & & 104.5/122	\\
00010381114	&	58657.66776	&	2019-06-23 16:01:34	&	2019-06-23 16:18:52	&	1038 &       0.9$^{+0.4}_{-0.3}$     &       1.1$^{+0.7}_{-0.6}$         &       4.5$^{+0.9}_{-0.8}$     &  & & 123.8/170	\\	
00010381115	&	58659.60588	&	2019-06-25 14:32:28	&	2019-06-25 14:47:53	&	925	&       0.8$^{+0.3}_{-0.3}$     &       0.7$^{+0.5}_{-0.5}$         &       30.0$^{+5.0}_{-5.0}$     &  & & 198.0/217	\\
00010381116	&	58661.46622	&	2019-06-27 11:11:21	&	2019-06-27 11:24:53	&	634	&       1.0$^{+0.9}_{-0.6}$     &       0.7$^{+1.1}_{-1.0}$         &       4.0$^{+1.2}_{-0.9}$     &  & & 94.6/96	\\
00010381117	&	58663.65581	&	2019-06-29 15:44:22	&	2019-06-29 15:59:52	&	930	&       1.5$^{+0.9}_{-0.6}$     &       1.4$^{+1.0}_{-0.9}$         &       3.9$^{+1.0}_{-0.8}$     &  & & 91.9/123	\\
00010381118	&	58665.37906	&	2019-07-01 09:05:50	&	2019-07-01 09:19:53	&	\multirow{2}{*}{1775}   &       \multirow{2}{*}{1.7$^{+1.4}_{-0.9}$}    &       \multirow{2}{*}{0.9$^{+1.3}_{-1.1}$}  &      \multirow{2}{*}{1.5$^{+0.5}_{-0.4}$}    & & &  \multirow{2}{*}{75.3/109}  \\	
00010381119	&	58667.63913	&	2019-07-03 15:20:20	&	2019-07-03 15:35:53	&			& & & &  & & 	\\
00010381120	&	58669.23266	&	2019-07-05 05:35:02	&	2019-07-05 05:49:52	&	890	&       1.0$^{+1.0}_{-0.6}$     &       0.6$^{+1.3}_{-1.1}$         &       1.8$^{+0.7}_{-0.5}$     &  & & 45.8/68	\\
00010381121	&	58671.35797	&	2019-07-07 08:35:29	&	2019-07-07 08:49:54	&	865	&       2.5$^{+1.3}_{-1.0}$     &       2.2$^{+1.4}_{-1.2}$         &       2.5$^{+0.8}_{-0.6}$     &  & & 90.3/99	\\
00010381122	&	58673.21462	&	2019-07-09 05:09:02	&	2019-07-09 05:24:53	&	948	&       1.2$^{+0.7}_{-0.6}$     &       1.0$^{+1.0}_{-0.9}$         &       3.7$^{+1.0}_{-0.8}$     &  & & 123.7/131	\\
00010381123	&	58675.20600	&	2019-07-11 04:56:38	&	2019-07-11 05:12:53	&	975	&       1.6$^{+1.1}_{-0.7}$     &       1.0$^{+1.2}_{-1.0}$         &       5.0$^{+1.4}_{-1.1}$     &  & & 83.1/116	\\
00010381124	&	58681.18158	&	2019-07-17 04:21:28	&	2019-07-17 04:37:53	&	985	&       1.3$^{+0.7}_{-0.5}$     &       1.5$^{+1.0}_{-0.9}$         &       2.6$^{+0.7}_{-0.5}$     &  & & 84.2/122	\\
00010381125	&	58683.17534	&	2019-07-19 04:12:29	&	2019-07-19 04:26:54	&	865	&       1.4$^{+0.6}_{-0.5}$     &       1.8$^{+0.9}_{-0.8}$         &       3.0$^{+0.7}_{-0.6}$     &  & & 125.8/142	\\
00010381126	&	58685.10042	&	2019-07-21 02:24:35	&	2019-07-21 02:35:57	&	682	&       1.3$^{+1.0}_{-0.7}$     &       1.0$^{+1.3}_{-1.1}$         &       3.1$^{+1.1}_{-0.8}$     &  & & 64.2/96	\\
00010381127	&	58687.29223	&	2019-07-23 07:00:48	&	2019-07-23 07:16:53	&	965	&       1.2$^{+0.7}_{-0.6}$     &       1.0$^{+1.0}_{-0.9}$         &       3.7$^{+1.0}_{-0.8}$     &  & & 123.7/131	\\
00010381128	&	58689.33925	&	2019-07-25 08:08:30	&	2019-07-25 09:54:53	&	820	&       4.9$^{+2.8}_{-2.1}$     &       2.9$^{+2.1}_{-1.7}$         &       2.0$^{+0.7}_{-0.5}$     &  & & 57.8/90	\\
00010381129	&	58691.41148	&	2019-07-27 09:52:31	&	2019-07-27 10:07:54	&	923	&       5.5$^{+2.4}_{-1.9}$     &       2.3$^{+1.5}_{-1.4}$         &       13.5$^{+3.6}_{-2.7}$     &  & & 123.5/130	\\
00010381130	&	58693.12654	&	2019-07-29 03:02:12	&	2019-07-29 03:04:53	&	\multirow{2}{*}{1098}   &       \multirow{2}{*}{1.3$^{+0.8}_{-0.6}$}    &       \multirow{2}{*}{0.7$^{+0.9}_{-0.8}$}  &      \multirow{2}{*}{3.5$^{+0.9}_{-0.7}$}    & & &  \multirow{2}{*}{104.0/136}  \\	
00010381131	&	58695.85434	&	2019-07-31 20:30:15	&	2019-07-31 20:45:52	&			& & & &  & & 	\\
00010381132	&	58697.91528	&	2019-08-02 21:57:59	&	2019-08-02 22:13:52	&	953	&       1.6$^{+1.1}_{-0.8}$     &       0.1$^{+1.0}_{-0.9}$         &       4.3$^{+1.1}_{-0.9}$     &  & & 120.8/125	\\
00010381133	&	58699.44008	&	2019-08-04 10:33:43	&	2019-08-04 10:51:53	&	1091	&       1.7$^{+0.7}_{-0.6}$     &       1.3$^{+0.8}_{-0.7}$         &       4.0$^{+0.8}_{-0.7}$     &  & & 171.0/175	\\
00010381134	&	58701.36509	&	2019-08-06 08:45:43	&	2019-08-06 09:03:54	&	1091 	&       1.5$^{+0.8}_{-0.6}$     &       1.9$^{+1.1}_{-1.0}$         &       2.1$^{+0.6}_{-0.4}$     &  & & 99.7/119	\\
00010381135	&	58703.43123	&	2019-08-08 10:20:58	&	2019-08-08 10:35:53	&	895		&       1.8$^{+0.8}_{-0.7}$     &       1.7$^{+1.0}_{-0.9}$         &       3.8$^{+0.9}_{-0.7}$     &  & & 125.7/139 \\
00010381136	&	58707.74719	&	2019-08-12 17:55:57	&	2019-08-12 18:09:52	&	\multirow{2}{*}{1810}   &       \multirow{2}{*}{0.9$^{+1.1}_{-0.7}$}    &       \multirow{2}{*}{-0.2$^{+1.3}_{-1.1}$}  &      \multirow{2}{*}{1.9$^{+0.8}_{-0.5}$}    & & &  \multirow{2}{*}{97.8/100}  \\	
00010381137	&	58709.53237	&	2019-08-14 12:46:36	&	2019-08-14 13:02:52	&			& & & &  & & 	\\
00010381138	&	58711.19394	&	2019-08-16 04:39:16	&	2019-08-16 04:50:28	&		\multirow{2}{*}{1427}   &       \multirow{2}{*}{2.5$^{+1.3}_{-1.1}$}    &       \multirow{2}{*}{2.3$^{+1.3}_{-1.2}$}  &      \multirow{2}{*}{2.0$^{+0.5}_{-0.4}$}    & & &  \multirow{2}{*}{113.0/138}  \\
00010381139	&	58719.28919	&	2019-08-24 06:56:26	&	2019-08-24 07:09:00	&			& & & &  & & 	\\
00010381140	&	58721.08657	&	2019-08-26 02:04:39	&	2019-08-26 02:19:52	&	993 &       2.4$^{+1.2}_{-1.0}$     &       2.0$^{+1.3}_{-1.1}$         &       3.0$^{+0.9}_{-0.6}$     &  & & 102.1/114	\\
00010381142	&	58725.40668	&	2019-08-30 09:45:36	&	2019-08-30 10:00:51	&	915		&       1.7$^{+1.1}_{-0.8}$     &       0.6$^{+1.1}_{-0.9}$         &       5.2$^{+1.4}_{-1.1}$     &  & & 86.6/119	\\
00010381143	&	58727.27516	&	2019-09-01 06:36:13	&	2019-09-01 06:39:51	&	\multirow{2}{*}{2008}   &       \multirow{2}{*}{3.8$^{+2.0}_{-1.6}$}    &       \multirow{2}{*}{1.7$^{+0.5}_{-0.4}$}  &      \multirow{2}{*}{1.6$^{+0.5}_{-0.4}$}    & & &  \multirow{2}{*}{87.5/122}  \\
00010381144	&	58729.25917	&	2019-09-03 06:13:12	&	2019-09-03 06:27:52	&		& & & &  & & 	\\
00010381145	&	58731.24285	&	2019-09-05 05:49:42	&	2019-09-05 06:04:52	&		& & & &  & & 	\\
00010381146	&	58735.83353	&	2019-09-09 20:00:16	&	2019-09-09 20:14:54	&	878	&       2.1$^{+1.3}_{-0.9}$     &       1.3$^{+1.2}_{-1.0}$         &       4.1$^{+1.1}_{-0.9}$     &  & & 95.4/114 \\
00010381148	&	58739.15327	&	2019-09-13 03:40:42	&	2019-09-13 03:56:53	&	970	&       1.0$^{+2.9}_{-0.5}$     &       0.2$^{+1.8}_{-0.7}$         &       10.3$^{+2.2}_{-2.6}$     &  & & 155.9/165 \\
00010381149	&	58741.14633	&	2019-09-15 03:30:42	&	2019-09-15 03:47:53	&	\multirow{2}{*}{2016}   &       \multirow{2}{*}{1.9$^{+1.1}_{-0.9}$}    &       \multirow{2}{*}{1.0$^{+1.1}_{-1.0}$}  &      \multirow{2}{*}{2.0$^{+0.5}_{-0.4}$}    & & &  \multirow{2}{*}{132.1/148} 	\\
00010381150	&	58743.60379	&	2019-09-17 14:29:27	&	2019-09-17 14:45:52	&		& & & &  & & 	\\
00010381151	&	58745.26425	&	2019-09-19 06:20:31	&	2019-09-19 06:36:51	&		\multirow{2}{*}{1848}   &       \multirow{2}{*}{2.3$^{+1.3}_{-1.0}$}    &       \multirow{2}{*}{0.7$^{+1.0}_{-0.9}$}  &      \multirow{2}{*}{3.0$^{+0.6}_{-0.5}$}    & & &  \multirow{2}{*}{152.3/163}  \\
00010381152	&	58747.32460	&	2019-09-21 07:47:25	&	2019-09-21 08:01:53	&	 & & & &  & & 	\\
00010381154	&	58751.70606	&	2019-09-25 16:56:43	&	2019-09-25 17:12:54	&	970		&       1.6$^{+0.7}_{-0.6}$     &       2.0$^{+1.0}_{-0.9}$         &       4.4$^{+1.1}_{-0.8}$     &  & & 120.1/124	\\
00010381155	&	58753.23374	&	2019-09-27 05:36:34	&	2019-09-27 05:50:52	&	857		&       1.0$^{+0.4}_{-0.3}$     &       0.6$^{+0.5}_{-0.5}$         &       11.2$^{+1.7}_{-1.4}$     &  & & 213.4/280	\\
00010381156	&	58755.35575	&	2019-09-29 08:32:17	&	2019-09-29 08:50:52	&	1116 	&       1.3$^{+0.6}_{-0.5}$     &       1.3$^{+0.8}_{-0.7}$         &       7.1$^{+8.4}_{-2.0}$     &  & & 149.3/163		\\
00010381158	&	58763.32481	&	2019-10-07 07:47:43	&	2019-10-07 08:02:53	&	910 	&       1.5$^{+0.8}_{-0.6}$     &       1.5$^{+1.0}_{-0.9}$         &       7.5$^{+20.3}_{-2.9}$     &  & & 90.8/128		\\
00010381159	&	58765.18065	&	2019-10-09 04:19:46	&	2019-10-09 04:35:51	&		\multirow{2}{*}{1093}   &       \multirow{2}{*}{0.9$^{+0.5}_{-0.4}$}    &       \multirow{2}{*}{0.4$^{+0.7}_{-0.6}$}  &      \multirow{2}{*}{2.5$^{+0.5}_{-0.4}$}    & & &  \multirow{2}{*}{159.5/175}  \\
00010381160	&	58767.36462	&	2019-10-11 08:45:03	&	2019-10-11 09:00:53	&	& & & &  & & 	\\
00010381161	&	58769.49017	&	2019-10-13 11:45:50	&	2019-10-13 12:02:53	&	1023 	&       1.4$^{+0.6}_{-0.5}$     &       1.3$^{+0.8}_{-0.7}$         &       11.6$^{+2.5}_{-2.0}$     &  & & 153.5/175	\\
00010381162	&	58771.15377	&	2019-10-15 03:41:25	&	2019-10-15 03:58:53	&	1048	&       1.2$^{+0.7}_{-0.5}$     &       1.1$^{+0.9}_{-0.8}$         &       4.8$^{+1.2}_{-0.9}$     &  & & 115.8/151	 \\
00010381163	&	58773.00434	&	2019-10-17 00:06:14	&	2019-10-17 21:15:53	&	1048 	&       2.1$^{+0.9}_{-0.7}$     &       1.6$^{+0.9}_{-0.8}$         &       4.2$^{+0.8}_{-0.7}$     &  & & 169.9/187 \\
00010381164	&	58775.27495	&	2019-10-19 06:35:55	&	2019-10-19 06:51:53	&	958	 	&       2.0$^{+0.8}_{-0.7}$     &       2.1$^{+1.0}_{-0.9}$         &       5.0$^{+1.1}_{-0.9}$     &  & & 113.7/172\\
00010381165	&	58777.33271	&	2019-10-21 07:59:06	&	2019-10-21 08:14:54	&	948		&       1.4$^{+0.7}_{-0.5}$     &       1.3$^{+0.9}_{-0.8}$         &       6.8$^{+1.5}_{-1.2}$     &  & & 145.3/173 \\
  \noalign{\smallskip}
  \hline
  \end{tabular}
  \end{center}
  \end{table*}

We first extracted the source average count-rate for each observation in two energy bands, 0.5--4\,keV and 4--10\,keV, to compute the hardness-ratio (HR) reported also in Table~\ref{i17329:tab:swift_xrt_log}. We then extracted for each observation the average spectrum of the source and fit it with a simple absorbed power-law model (the same as used in Sect.~\ref{sec:sctx1}). We note that the XRT observations of \srcb\ are generally composed by 1 up to 3 snapshots with typical exposures of 300--1000\,s. Therefore, none of the XRT observations is suitable to look for pulsations; furthermore the energy dependence of the pulse profile is unlikely to bias the HR value computed for each observation. The results of these analysis are summarized in Fig.~\ref{fig:17329_xrt_lc}.

The energy spectral distribution of the source in the XRT energy band is generally well described by simple absorbed power-law model, but there are intervals of time where the fits to the spectra with such simple model leave evident residuals around the typical energy of the neutral iron line at $\sim$6.4~keV. During these intervals, the source is relatively faint and achieves a flux that is a factor of 5--10 lower than the surrounding time intervals. We also observe simultaneously during these intervals a flattening of the spectrum and a drop by a factor of few in the absorption column density, thus resembling what we usually observe in wind-fed systems during X-ray eclipses or obscuration events \citep[see, e.g.,][and references therein]{bozzo09}. Although all spectral results are reported in Table~\ref{i17329:tab:swift_xrt_log}, we also show a plot of the parameters as a function of time in Fig.~\ref{fig:17329_xrt_lc} to ease the visual inspection. As it can be appreciated for this figure, the intervals corresponding to the obscuration events are concentrated during the second observational period (58150--58396~MJD) and occur in the time intervals separating the different source flares observed (4 flares are apparent in total during this period). The time interval characterized by the iron line with the largest equivalent width ($\gtrsim$2~keV) is centered around $\sim$58240~MJD and it is also featuring the lowest recorded flux as well as one of the softest measured X-ray spectra. To illustrate the dramatic change in the source spectral properties between the emission during and outside the obscuration events, we show a comparison of typical spectra for the two cases in Fig.~\ref{fig:17329_iron}.

Although the uncertainties associated with the measurements of the iron line centroid energies in the different obscuration intervals in Table~\ref{i17329:tab:swift_xrt_log} is relatively large, we see also from Fig.~\ref{fig:17329_iron} a marginal evidence for a change of the centroid between 6.4~keV up to 6.7~keV. In order to improve the measurements of the iron line, we thus stacked together all data collected during the obscuration intervals and obtained the spectrum that we show in Fig.~\ref{fig:17329_xrt_obscuration}. As expected based on the findings above, the spectrum appears relatively flat, and there is a prominent feature around 6.4~keV emerging from the residuals if the spectrum is fit with a simple absorbed powerlaw (in this case the result would be largely unacceptable with $\chi^{2}/d.o.f.$=147.5/45). Adding a single thin Gaussian component (width fixed to zero)  improves the fit ($\chi^{2}/d.o.f.$=73.9/43) but leaves significant residuals in the energy range 6.7--7.0~keV. Adding a second Gaussian component centered at 6.7~keV further improves the fit down to $\chi^{2}/d.o.f.$=49.2/41. From this fit, we measured an absorption column density of $N_{\rm H}$=(2.8$^{+2.0}_{-1.4}$)$\times$10$^{22}$~cm$^{-2}$, a powerlaw photon index of $\Gamma$=-0.8$\pm$0.2, and a 0.5--10~keV flux of 1.2$\times10^{-11}$~erg cm$^{-2}$ s$^{-1}$  (not corrected for absorption). The measured centroid energies of the two iron lines in this case are 6.41$\pm$0.06~keV and 6.8$\pm$0.1~keV. The corresponding equivalent widths (EQWs) were 0.83$^{+0.17}_{-0.40}$~keV and 0.48$^{+0.12}_{-0.30}$~keV, respectively. Although the fit is formally acceptable, we still noticed residuals around 7~keV. We thus attempted to add a third Gaussian line and obtained a centroid energy for this line of 6.9$\pm$0.1~keV while all other parameters remained unchanged to within the uncertainties. The improvement of the fit with the third line is not statistically significant ($\chi^{2}/d.o.f.$=39.6/39), but it is interesting to note that the three lines correspond to the known predominant states of iron typically observed in wind-fed HMXBs \citep[neutron iron line at 6.4~keV, He-like iron line at 6.7~keV, and H-like iron line at 6.9~keV; see, e.g.,][and references therein]{fuerst11,liu18}. For completeness, we mention that the residuals around the iron line complex could also be alternatively fit ($\chi^{2}/d.o.f.$=40.1/42) with a single large Gaussian component (leaving the continuum parameters virtually unchanged). In this case, the centroid energy would be 6.52$\pm$0.06~keV and the line width 0.27$\pm$0.07~keV. We consider this fit unphysical as the centroid energy of the line would only be marginally compatible with the neutral iron value and a broadening of the line would only be expected in case of a fast rotating accretion disk around a weakly magnetized NS \citep[see, e.g.,][for a recent review]{salvo20} but not in a wind-fed young pulsar as that hosted in \srcb\ \citep[see the discussion in][]{torrejon10}.

\section{Discussion}
\label{sec:discussion}

We reported in this paper about the broad-band X-ray spectroscopy of two of the few known SyXBs, \srca, and \srcc.\ We have been looking specifically for the presence of CRSFs in their spectral energy distribution that could point toward the presence of strongly magnetized NSs in these systems and support the idea that they can be formed through the accretion induced collapse of a white dwarfs (a yet relatively poorly known NS formation channel, see Sect.~\ref{sec:intro}). Furthermore, we reported on our long-term monitoring campaign on the latest discovered SyXB, \srcb,\ carried out with the \swift/XRT up to nine months following the first detection of the source in the X-ray domain.

In the case of \srca,\ we exploited our simultaneous \chan\ and \nustar\ observation performed in 2020. The source turned out to be relatively faint, with an average flux that decreased about two orders of magnitude compared to the previous X-ray observations carried out in 2004 with \xmm.\ The slowly decreasing trend in the source flux is confirmed also by two \suzaku\ observations performed in 2014 where the source was not detected (the upper limit we provided is a factor of few higher than the flux values determined by both \chan\ and \nustar). Although the statistics of the combined \chan\ and \nustar\ data was relatively poor, we could shows that the source dimming over the past decades is accompanied by a moderate softening of the spectrum (the power-law photon index increased from $\sim$1.5 in 2004 to roughly $\sim$3.0 in 2020). No evidence of CRSF was found in the broad-band spectrum of the source. The slow decade-long dimming and softening of the source emission is not easy to interpret in the context of wind-fed SyXBs, but it is worth noticing that such behavior (accompanied by a local absorption column density constantly in excess of 10$^{23}$~cm$^{-2}$) was already reported for the other (yet poorly known) SyXB XTE\,J1743-363 \citep{smith12,bozzo13}.

For \srcc,\ we looked for the possible presence of CRSFs in the X-ray spectral energy distribution of the source by exploiting a public (but not yet published) \nustar\ observation of the source, as well as simultaneous XRT+BAT spectra collected during three outbursts of the source caught by \swift\ in 2014 and 2015. When the broad-band spectrum obtained from both the FPMA and FPMB on-board \nustar\ is fit with a simple cut-off power-law model, significant residuals are found especially around 15~keV. These residuals suggested the presence of a broad absorption feature, similarly to what is usually expected for a CRSF. Interestingly, the FPMA and FPMB spectra also showed some marginal evidence for the first harmonic of this feature due to additional residuals around 30~keV. If our interpretation is correct, the NS hosted in \srcc\ could be endowed with a magnetic field strength of $ \sim$1.4$\times$10$^{12}$~G \citep[see, e.g.][for the conversion of the CRSF centroid energy line into the NS magnetic field strength]{staubert19}. This is the first claim of a CRSFs in the X-ray spectrum of \srcc.\ We note that past investigations on the X-ray broad-band spectral energy distribution of the source were carried out (to the best of our knowledge) only by \citet{masetti02} and \citet{nagae08}. These authors exploited mainly \beppo,\ and \suzaku\ data where the energy range 10-30~keV is poorly covered (or missing) due to the limited overlap of the operational energies of the available instruments. \citet{masetti02} also reported on a \rxte/PCA spectrum of the source extending continuously from 3 to 20~keV. Also in this case, no CRSF was reported. We remark, however, that the fundamental line identified in our \nustar\ spectra would sit right at the rim of the PCA energy coverage of that spectrum and thus might have easily gone undetected in the fit. There is also the possibility that the CRSF was not visible during the PCA observation, as we known that CRSF might be dependent on the viewing direction of the observer compared to the magnetic field orientation and could change along the pulse period or at different luminosity levels \citep[impacting, e.g., the shape of the accretion column where the CRSFs are most likely originating; see, e.g.,][and references therein]{kre20}.

However, it must be remarked that the present statistics of the \nustar\ data does not allow us to firmly exclude alternative models to describe the broad-band spectral energy distribution of \srcc.\ The FPM spectra could be equivalently well described using the addition of a black-body component to the cut-off power-law in place of the two absorption features. The presence of a similar thermal component would not be unexpected, as it was previously reported for this source by \citet{nucita14} while analyzing \xmm\ and \swift\/XRT data outside the outbursts. If we compare the results from these authors with those in Table~\ref{tab:4u_spectral}, we note that the black-body component required by the \nustar\ data is significantly hotter (1.4~keV vs. the previously reported 0.7-1.0~keV) and would be dominating a large portion of the XRT energy coverage. Curiously enough, the broad-band XRT$+$BAT spectra that we obtained by analyzing the source behavior during three different outbursts did not show the presence of any thermal component. Although the energy coverage and statistics of all XRT+BAT data did not allow us to investigate in details the presence of the CRSF at $\sim$16~keV and its harmonic, we discussed in Sect.~\ref{sec:4u1700} that the broad-band fits would be compatible with the (undetected) presence of these features, if their parameters are frozen to those measured by \nustar.\

We thus conclude that at present it is not possible to firmly confirm or reject the presence of CRSFs in the broad-band X-ray spectrum of \srcc.\  Deeper observations of this system with \nustar,\ featuring the uniquely required combination of sensitivity and large energy coverage, are needed to understand if \srcc\ is the second SyXB to host a strongly magnetized NS and evaluate further the impact of these results for the formation channels of extreme compact objects from accreting aged white dwarfs.

\srcb\ was the first SyXB with a detected CRSF, leading to the identification in this system of a young NS endowed with a magnetic field strength of $\sim$2$\times$10$^{12}$~G (see Sect~\ref{sec:intro}). The observations collected during our long-term XRT monitoring showed that the source never turned off as an X-ray emitter after its initial detection and remained active at a virtually constant average flux level of few times 10$^{-11}$~$\ferg$. It also underwent sporadic flares lasting a few days at the most and achieving a total dynamic range in the X-ray flux of about $\sim$30 (see Fig.~\ref{fig:17329_xrt_lc} and Table~\ref{i17329:tab:swift_xrt_log}).

On one hand, the findings of the monitoring program corroborated the idea proposed by \citet{bozzo18} that \srcb\ turned on in 2017 for the first time as a SyXB, beginning effectively to be a variable but persistent source as all other objects in this class. On the other hand, the reason for the onset of such X-ray activity remains unclear. \citet{bozzo18} proposed that the red giant might have gone through some thermal pulse phase that had enhanced the mass loss rate toward the NS and so triggered a sufficiently intense accretion to be detectable for the first time in X-rays (explaining also the brightening of the optical star in the R-band). We can investigate here a slightly different possibility.

The flares and in general the prominent X-ray variability of the source revealed by our monitoring is commonly observed in SyXBs and usually ascribed to the presence of structures in the wind of the red giant companion \citep[see, e.g.,][and references therein]{yungelson19}. It is well known that the slow ($\varv_{\rm wind}\sim 10$\,{\rm km}\,s$^{-1}$) dusty winds of cool evolved stars are strongly inhomogeneous and contain large scale structures such as clumps and shells \citep[for a recent review see][]{Decin2021}. The structures have been directly resolved in the atmospheres of some nearby M-type giants using imaging and interferometric observations. \citet{Adam2019} show that the clumpy dusty clouds are present already within a few stellar radii in the wind of IK\,Tau, while the asymmetric dust emission extends to outer wind regions. The clumpy structures have typical density enhancements of a factor of $\sim$3 compared to the surrounding wind. \citet{Adam2019} measured the extent of two huge clumps which have a size of $\approx 80$\,mas.  At the distance $260$\,pc to the star, the measured clump size corresponds to $\approx 10^{9}$\,km. Assuming a typical orbital velocity of $10$\,km\,s$^{-1}$ for a NS around a red giant with a period of several years, it would take the compact object about 1\,yr to cross a similarly extended structure. It is thus possible that the enhancement in the mass accretion rate that was needed by \srcb\ to finally turn on as a SyXB has been caused by the NS interaction with one (or more) large stellar wind clump. If the orbit of the system is as large (or larger) than that measured for \srcc\ \citep{hinkle19}, then we might conclude that in the past the NS has never being in contact with similarly large structures and thus no X-ray emission could ever be detected before 2017. According to this scenario, it is possible that \srcb\ will stop shining in X-rays over the time scale of one to few years, as soon as the orbit of the compact object does no longer intersect a massive stellar wind clump.

The XRT monitoring of \srcb\ also revealed the presence of intriguing obscuration events, lasting a few days (see Fig.~\ref{fig:17329_iron} and \ref{fig:17329_xrt_obscuration}). The spectral properties observed during these events are remarkably similar to those commonly recorded during X-ray eclipses. However, it is unlikely that the NS in \srcb\ was eclipsed by its giant companion during the XRT observations. The main argument against this possibility is that the obscuration events revealed by XRT have different durations and they repeat irregularly for 5 times in less than $\sim$100~days (to be compared to the typical orbital period of SyXBs of the order of years). A more likely possibility, already explored in the case of supergiant X-ray binaries where the NS is accreting from the clumpy wind of its massive companion, is that the NS in \srcb\ was temporarily obscured by a stellar wind structure located along the line of sight to the observer \citep[see, e.g.,][and references therein]{bozzo11,nunez17}. As we considered above that a single massive clump could be at the origin of the onset of the X-ray activity from \srcb,\ we can assume that the obscuration events are most likely triggered by the presence of sub-structures within the large clump or in the surroundings. There are indications, indeed, that atmospheric structures of red giants could change on the time scale of weeks as confirmed by both observations and hydrodynamic models \citep[see][and references therein]{Hofner2019}. Furthermore, their stellar winds are permeated by shock fronts which are induced by stellar pulsations and/or convection \citep[see, e.g.,][]{Perrin2020}. The presence of a NS embedded in the wind further perturbs the atmospheric morphology with its orbital motion as well as with the effect of wind ionization by the X-ray radiation \citep[see, e.g.,][and references therein]{bozzo21}. We can thus envision the possibility of the smaller scale structures being present in the wind or within the clump to induce the temporary obscuration events, as well as the short flares observed by XRT. The detection of emission lines corresponding to (at least)  He-like and H-like iron further supports this scenario as the ionization of the stellar wind material is most likely occurring in the vicinity of the NS due to the X-ray radiation.

\section*{Data availability}
The data underlying this article are publicly available from the \xmm,\ \nustar,\ \chan,\ and \swift\ archives and processed with publicly available software.

\section*{Acknowledgements}
We thank the anonymous referee for swift comments that helped us improve the paper.
The \swift\ data of our monitoring campaigns on \srcb\ were obtained through contract ASI-INAF I/004/11/5 (PI P.\ Romano). PR acknowledges financial contribution from contract ASI-INAF ASI-INAF I/037/12/0. 
This work made use of data supplied by the UK Swift Science Data Centre at the
University of Leicester \citep[see][]{2007A&A...469..379E,2009MNRAS.397.1177E}.

\bibliography{bib.bib}{}
\bibliographystyle{mnras}

\end{document}